\documentclass[12pt]{article}
\usepackage{amsmath}
\usepackage{graphicx,psfrag,epsf}
\usepackage{enumerate}
\usepackage{natbib}
\usepackage{url} % not crucial - just used below for the URL 
\usepackage{amsmath, amsthm, amssymb}
\usepackage{graphicx}
\usepackage{subfig}
\usepackage{array}
\newcommand{\blind}{0}

\theoremstyle{plain}
\newtheorem{thm}{\protect\theoremname}
  \theoremstyle{plain}
  \newtheorem{prop}[thm]{\protect\propositionname}
 \providecommand{\propositionname}{Proposition}
\providecommand{\theoremname}{Theorem}
\DeclareMathOperator*{\argmax}{arg\,max}

\begin{document}

\def\spacingset#1{\renewcommand{\baselinestretch}%
{#1}\small\normalsize} \spacingset{1}

%%%%%%%%%%%%%%%%%%%%%%%%%%%%%%%%%%%%%%%%%%%%%%%%%%%%%%%%%%%%%%%%%%%%%%%%%%%%%%

\if0\blind
{
  \title{\bf Image-Based Prognostics Using Penalized Tensor Regression}
  \author{Xiaolei Fang, Kamran Paynabar, Nagi Gebraeel\hspace{.2cm}\\
   H. Milton Stewart School of Industrial and Systems Engineering\\
Georgia Institute of Technology\\}
  \maketitle
} \fi

\if1\blind
{
  \bigskip
  \bigskip
  \bigskip
  \begin{center}
    {\LARGE\bf  Image-Based Prognostics Using Penalized Tensor Regression}
\end{center}
  \medskip
} \fi

\bigskip
\begin{abstract}
This paper proposes a new methodology to predict and update the residual
useful lifetime of a system using a sequence of degradation images.
The methodology integrates tensor linear algebra with traditional
location-scale regression widely used in reliability and prognosis.
To address the high dimensionality challenge, the degradation image
streams are first projected to a low-dimensional tensor subspace that
is able to preserve their information. Next, the projected image tensors
are regressed against time-to-failure via penalized location-scale
tensor regression. The coefficient tensor is then decomposed using
CANDECOMP/PARAFAC (CP) and Tucker decompositions, which enables parameter
estimation in a high-dimensional setting. Two optimization algorithms
with a global convergence property are developed for model estimation.
The effectiveness of our models is validated using a simulated dataset
and infrared degradation image streams from a rotating machinery. 
\end{abstract}

\noindent%
\vfill
%\hfill {\tiny technometrics tex template (do not remove)}

\newpage
\spacingset{1.45} % DON'T change the spacing!
\section{Introduction}

\label{sec:introduction}

Imaging is one of the fastest growing technologies for condition monitoring
and industrial asset management. Relative to most sensing techniques,
industrial imaging devices are easier to use because they are generally
noncontact and do not require permanent installation or fixturing.
Image data also contains rich information about the object being monitored.
Some examples of industrial imaging technologies include infrared
images used to measure temperature distributions of equipment and
components \citep{Bagavathiappan2013}, charge-coupled device (CCD)
images which capture surface quality information (e.g., cracks) of
products \citep{Neogi2014}, and others. Image data has been extensively
used for process monitoring and diagnostics. For instance, infrared
images have been successfully used for civil structures monitoring
\citep{Meola2007}, machinery inspection \citep{Seo2011}, fatigue
damage evaluation \citep{Pastor2008} and electronic printed circuit
board (PCB) monitoring \citep{Vellvehi2011}. In steel industry, CCD
cameras have been utilized for product surface inspection \citep{Neogi2014},
while video cameras have been used to monitor the shape and color
of furnace flames to control quality of steel tubes \citep{Yan2015}.
This paper expands the utilization of image data by proposing an image-based
prognostic modeling framework that uses degradation-based image streams
to predict remaining lifetime.

Numerous prognostic methodologies have been developed in the literature.
Examples of some modeling approaches include random coefficients models
\citep{Gebraeel,Ye2014}, models that utilize the Brownian motion
process \citep{Ye2015,Chen2015} and gamma process \citep{Shu2015,Zhang2015},
and models based on functional data analysis \citep{Fang2015,Zhou2014}.
These approaches are well-suited for time-series signals, but it is
not clear how they can be extended to model image streams. One of
the key challenges in modeling image data revolves around the analytical
and computational complexities associated with characterizing high
dimensional data. High dimensionality arises from the fact that a
single image stream consists of a large sequence of images (observed
across the life cycle of an equipment) coupled with the large numbers
of pixels embedded in each image. Additional challenges are related
to the complex \textit{spatial-temporal structures} inherent in the
data streams. Pixels are spatially correlated within a single image
and temporally correlated across sequential images. In recent work
\citep{Liu2016}, degradation image streams were modeled as a spatio-temporal
process. Although spatio-temporal models have been widely used to
model data with complex spatial and temporal correlation structures
\citep{Cressie2015}, they are not necessarily accurate for long-term
predictions necessary to our type of prognostic application. Most
importantly, a key limitation of spatio-temporal models is that they
require a pre-set failure threshold, which is usually hard to define
for degradation image streams.

This paper proposes a tensor-based regression framework that utilizes
degradation image streams to predict remaining useful life (RUL),
and provide advance warning of impending failures of industrial assets.
Specifically, we build a (log)-location-scale (LLS) tensor regression
model in which the time-to-failure (TTF) is treated as the response
and degradation image streams as covariates. LLS regression has been
widely used in reliability and survival analysis \citep{Doray94}
because it provides a flexible framework capable of modeling a variety
of TTF distributions such as \textit{(log)normal, (log)logistic, smallest
extreme value (SEV)}, \textit{Weibull}, etc. To model the \textit{spatio-temporal
structure} of degradation image streams, the regression model treats
each image stream as a \textit{tensor}. A tensor is defined as a \textit{multi-dimensional
array}\textendash a one-order tensor is a vector, a two-order tensor
is a matrix, and objects of order three or higher are called high-order
tensors. More details about tensor theory and applications can be
found in a survey paper by \citet{Kolda2009}. A degradation image
stream constitutes a three-order tensor in which the first two dimensions
capture the spatial structure of a single image whereas the third
dimension is used to model the temporal structure of the image stream. One of the key benefits
of modeling a degradation image stream as a tensor is that tensors
maintain the \textit{spatio-temporal structure} within and between
images which allows for a relatively accurate RUL prediction model. In this paper, \emph{degradation image stream(s)} and \emph{degradation
tensor(s)} are used exchangeably hereafter. 

The high dimensionality of degradation image streams poses significant
computational challenges, especially ones related to parameter estimation.
For example, a tensor-regression model for a degradation image stream
consisting of 50 images each with $20\times20$ pixels generates a
three-order tensor-coefficient consisting of 20,000 elements that
need to be estimated. In an effort to improve model computations,
we develop two estimation methods that integrate dimensionality
reduction and tensor decomposition. Dimensionality reduction is used
as the first step for both estimation methods as it helps reduce the
number of parameters. Degradation tensors are projected to a low-dimensional
tensor subspace that preserves most of their information. This is
achieved using a multilinear dimension reduction technique, such as
multilinear principal component analysis (MPCA) \citep{Lu2008}. We
utilize the fact that essential information embedded in high-dimensional
tensors can be captured in a low-dimensional tensor subspace. Next,
the tensor-coefficients corresponding to the projected degradation
tensors are decomposed using two popular tensor decomposition approaches
namely, CANDECOMP/PARAFAC (CP) \citep{Carroll1970} and Tucker \citep{Tucker1966}.
The CP approach decomposes a high-dimensional coefficient tensor as
a product of several low-rank basis matrices. In contrast, the Tucker
approach expresses the tensor-coefficient as a product of a low-dimensional
core tensor and several factor matrices. Therefore, instead of estimating
the tensor-coefficient, we only estimate its corresponding core tensors
and factor/basis matrices, which significantly reduces the computational
complexity and the required sample size. Block relaxation algorithms
are also developed for model estimation with guaranteed global convergence
to a stationary point.

The remainder of the paper is organized as follows. Section \ref{sec:pre}
provides an overview of the basic notations and definitions in multilinear
algebra. Section \ref{sec:reg} presents the degradation and prognostic
modeling framework. Section \ref{sec:cp} and \ref{sec:tucker} discusses
the estimation algorithm based on CP decomposition and Tucker decomposition,
respectively. In Section \ref{sec:updating}, we discuss the RUL prediction
and realtime updating. The effectiveness of our model is validated
using a numerical study in Section \ref{sec:simulation} and real
degradation image streams from a rotating machinery in Section \ref{sec:case}.
Finally, Section \ref{sec:conclusion} is devoted to concluding remarks.

\section{Preliminaries}

\label{sec:pre} This section presents some basic notations, definitions
and operators in multilinear algebra and tensor analysis that are
used throughout the paper. Scalars are denoted by lowercase letters,
e.g., $b$, vectors are denoted by lowercase boldface letters, e.g.,
$\boldsymbol{b}$, matrices are denoted by uppercase boldface letters,
e.g., $\boldsymbol{B}$, and tensors are denoted by calligraphic letters,
e.g., $\mathcal{B}$. The \textit{order} of a tensor is the number
of modes, also known as \textit{way}. For example, the order of vectors
and matrices are 1 and 2, respectively. A $D$-order tensor is denoted
by $\mathcal{B}\in\mathbb{R}^{I_{1}\times\cdots\times I_{D}}$, where
$I_{d}$ for $d=1,\ldots,D$ represents the dimension of the $d$-mode
of $\mathcal{B}$. The $(i_{1},i_{2},\ldots,i_{D})$-th component
of $\mathcal{B}$ is denoted by $b_{i_{1},i_{2},\ldots,i_{D}}$. A
\textit{fiber} of $\mathcal{B}$ is a vector obtained by fixing all
indices of $\mathcal{B}$ but one. A \textit{vectorization} of $\mathcal{B}$,
denoted by ${vec}({\mathcal{B}})$, is obtained by stacking all mode-$1$
fibers of ${\mathcal{B}}$. The mode-$d$ \textit{matricization} of
${\mathcal{B}}$, denoted by $\boldsymbol{B}_{(d)}$, is a matrix
whose columns are mode-$d$ fibers of ${\mathcal{B}}$ in the lexicographical
order. The mode-$d$ product of a tensor ${\mathcal{B}}\in\mathbb{R}^{I_{1}\times\cdots\times I_{D}}$
with a matrix $A\in\mathbb{R}^{J\times I_{d}}$, denoted by $({\mathcal{B}}\times_{d}A)$,
is a tensor whose component is $({\mathcal{B}}\times_{d}A)_{i_{1},\ldots,i_{d-1},j_{d},i_{d+1}\ldots,i_{D}}=\sum_{i_{d}=1}^{I_{d}}b_{i_{1},i_{2},\ldots,i_{D}}a_{j,i_{d}}$.
The \textit{inner product} of two tensors $\mathcal{B}\in\mathbb{R}^{I_{1}\times\cdots\times I_{D}},\mathcal{S}\in\mathbb{R}^{I_{1}\times\cdots\times I_{D}}$
is denoted by $\langle\mathcal{B},\mathcal{S}\rangle=\sum_{i_{1},\ldots,i_{D}}b_{i_{1},\ldots,i_{D}}s_{i_{1},\ldots,i_{D}}$.
A rank-one tensor $\mathcal{B}\in\mathbb{R}^{I_{1}\times\cdots\times I_{D}}$
can be represented by outer products of vectors, i.e., $\mathcal{B}=\boldsymbol{b}_{1}\circ\boldsymbol{b}_{2}\circ\cdots\circ\boldsymbol{b}_{D}$,
where $\boldsymbol{b}_{d}$ is an $I_{d}$-dimension vector and ``$\circ$\char`\"{}
is the \textit{outer product} operator. The \textit{Kronecker product}
of two matrices $\boldsymbol{A}\in\mathbb{R}^{m\times n},\boldsymbol{B}\in\mathbb{R}^{p\times q}$,
denoted by $\boldsymbol{A}\otimes\boldsymbol{B}$ is an $mp\times nq$
block matrix defined by

\[
M=\left(\begin{smallmatrix}a_{11}\boldsymbol{B} & \ldots & a_{1n}\boldsymbol{B}\\
\vdots & \ddots & \vdots\\
a_{m1}\boldsymbol{B} & \ldots & a_{mn}\boldsymbol{B}
\end{smallmatrix}\right).
\]

\noindent The \textit{Khatri-Rao} product of two matrices $\boldsymbol{A}\in\mathbb{R}^{m\times r},\boldsymbol{B}\in\mathbb{R}^{p\times r}$,
denoted by $\boldsymbol{A}\odot\boldsymbol{B}$, is a $mp\times r$
matrix defined by $\left[\boldsymbol{a}_{1}\otimes\boldsymbol{b}_{1}\;\;\boldsymbol{a}_{2}\otimes\boldsymbol{b}_{2}\;\;\cdots\;\;\boldsymbol{a}_{r}\otimes\boldsymbol{b}_{r}\right]$,
where $\boldsymbol{a}_{i}\in\mathbb{R}^{m\times1}$, and $\boldsymbol{b}_{i}\in\mathbb{R}^{p\times1}$
for $i=1,\ldots,r$.

\section{Prognostic Modeling Using Degradation Tensors}

\label{sec:reg} This paper considers engineering systems with degradation
process that can be represented by tensors, e.g., degradation image
streams or profiles. The underlying premise of our prognostic modeling
framework rests on using LLS regression to model TTF as a function
of degradation tensors. One of the main challenges in fitting such
regression models is the high-dimensionality of data which makes coefficients
estimation intractable. To address this issue, we use the fact that
the essential information of high-dimensional data is often embedded
in a low-dimensional subspace. Specifically, we project degradation
and coefficient tensors onto a low-dimensional tensor subspace that
preserves their inherent information.

To further reduce the number of estimated parameters, coefficient
tensors are decomposed using two widely used tensor decomposition
techniques, CP and Tucker. The CP decomposition expresses a high-dimensional
coefficient tensor as a product of several smaller sized basis matrices \citep{Carroll1970}. Tucker decomposition, however, expresses a high-dimensional
coefficient tensor as a product of a low-dimensional core tensor and
several factor matrices \citep{Tucker1966}. Thus, instead of estimating
the coefficient tensor, we only need to estimate its corresponding
core tensors and factor/basis matrices, which significantly helps
reduce the computational complexity and the required sample for estimation.
The parameters of the reduced LLS regression model are estimated using
the maximum likelihood (ML) approach. To obtain the ML estimates,
we propose  optimization algorithms for CP-based and Tucker-based
methods . The optimization algorithms are based on the block relaxation
method \citep{De1994,Lange2010}, which alternately updates one block of the
parameters while keeping other parameters fixed. Finally, the estimated
LLS regression is used to predict and update the RUL of a functioning
system. In the following, the details of the proposed methodology
is presented.

Our framework is applicable in settings that have a historical dataset
of degradation image streams (i.e., degradation tensor) for a sample
of units with corresponding TTFs. Let $N$ denote the number of units
that make up the historical (training) sample. Let $\mathcal{S}_{i}\in\mathbb{R}^{I_{1}\times\cdots\times I_{D}}$,
for $i=1,\ldots,N$, denote the degradation tensor and $\tilde{y}_{i}$
represent the TTF. The following LLS regression model expresses the
TTF as a function of a degradation tensor:

\begin{equation}
y_{i}=\alpha+\langle\mathcal{B},\mathcal{S}_{i}\rangle+\sigma\epsilon_{i}\label{eq:treg}
\end{equation}

\noindent where $y_{i}=\tilde{y}_{i}$ for a location-scale model
and $y_{i}=\ln(\tilde{y}_{i})$ for a log-location-scale model, the
scalar $\alpha$ is the intercept of the regression model, and $\mathcal{B}\in\mathbb{R}^{I_{1}\times\cdots\times I_{D}}$
is the tensor of regression coefficients. $\alpha+\langle\mathcal{B},\mathcal{S}_{i}\rangle$
is known as the location parameter and $\sigma$ is the scale parameter.
Similar to common LLS regression models \citep{Doray94}, we assume
that only the location parameter is a function of the covariates,
i.e., the degradation tensor. The term $\epsilon_{i}$ is the random
noise term with a standard location-scale density $f(\epsilon)$.
For example, $f(\epsilon)=\exp(\epsilon-\exp(\epsilon))$ for SEV
distribution, $f(\epsilon)=\exp(\epsilon)/(1+\exp(\epsilon))^{2}$
for logistic distribution, and $f(\epsilon)=1/\sqrt{2\pi}\exp(-\epsilon^{2}/2)$
for normal distribution. Consequently, $y_{i}$ has a density in the
form of $\frac{1}{\sigma}f\left(\frac{y_{i}-\alpha-\langle\mathcal{B},\mathcal{S}_{i}\rangle}{\sigma}\right)$.

The number of parameters in Model (\ref{eq:treg}) is given by $2+\prod_{d=1}^{D}I_{d}$.
Recall that $I_{d}$ for $d=1,\ldots,D$ represents the dimension
of the $d$-mode of $\mathcal{B}$. If we consider a simple example
of an image stream constituting 100 images of size $40\times50$,
i.e., $\mathcal{S}_{i}$ is a 3-order tensor in $\mathbb{R}^{40\times50\times100}$,
the number of parameters to be estimated will be quite large: $200,002=2+40\times50\times100$.
To reduce the number of parameters, as mentioned earlier, we project
the degradation tensors and the coefficient tensor onto a low-dimensional
tensor subspace that captures the relevant information of the degradation
tensors. The following proposition shows that by using multilinear
dimension reduction techniques, we can significantly reduce the dimensions
of the coefficient tensor without significant loss of information.
\begin{prop}
\label{prop:dim}Suppose $\{\mathcal{S}_{i}\}_{i=1}^{N}$ can be expanded
by $\mathcal{S}_{i}=\tilde{\mathcal{S}}_{i}\times_{1}\boldsymbol{U}_{1}\times_{2}\boldsymbol{U}_{2}\times\cdots\times_{D}\boldsymbol{U}_{D}$, where $\tilde{\mathcal{S}}_{i}\in\mathbb{R}^{P_{1}\times\cdots\times P_{D}}$ is a low-dimensional tensor and matrices $\boldsymbol{U}_{d}\in\mathbb{R}^{P_{d}\times I_{d}}$, $\boldsymbol{U}_{d}^{\top}\boldsymbol{U}_{d}=\boldsymbol{I}_{I_{d}}$, $P_{d}<I_{d}$, $d=1,\ldots, D$. If the coefficient
tensor, $\mathcal{B}$, is projected onto the tensor subspace spanned
by $\{\boldsymbol{U}_{1},\ldots,\boldsymbol{U}_{D}\}$, i.e., $\tilde{\mathcal{B}}=\mathcal{B}\times_{1}\boldsymbol{U}_{1}^{\top}\times_{2}\boldsymbol{U}_{2}^{\top}\times\cdots\times_{D}\boldsymbol{U}_{D}^{\top}$,
where $\tilde{\mathcal{B}}$ is the projected coefficient tensor,
then $\langle\mathcal{B},\mathcal{S}_{i}\rangle=\langle\tilde{\mathcal{B}},\tilde{\mathcal{S}}_{i}\rangle$. 
\end{prop}
The proof of Proposition \ref{prop:dim} is given in Appendix \ref{app:Proof-for-Propositiondim}.
Proposition \ref{prop:dim} implies that the original high-dimensional
tensors, (i.e., $\mathcal{B}$ and $\mathcal{S}$) and their corresponding
low-rank projections (i.e., $\tilde{\mathcal{B}}$ and $\tilde{\mathcal{S}}_{i}$)
result in similar estimates of the location parameter. Using Proposition
\ref{prop:dim}, we can re-express Equation (\ref{eq:treg}) as follows:

\begin{equation}
y_{i}=\alpha+\langle\tilde{\mathcal{B}},\tilde{\mathcal{S}}_{i}\rangle+\sigma\epsilon_{i}.\label{eq:treg2}
\end{equation}

The low-dimensional tensor space defined by factor matrices $\boldsymbol{U}_{d}\in\mathbb{R}^{P_{d}\times I_{d}}$
can be obtained by applying multilinear dimension reduction techniques,
such as multilinear principal component analysis (MPCA) \citep{Lu2008},
on the training degradation tensor, $\{\mathcal{S}_{i}\}_{i=1}^{N}$.
The objective of MPCA is to find a set of orthogonal factor matrices
$\{\boldsymbol{U}_{d}\in\mathbb{R}^{P_{d}\times I_{d}},\boldsymbol{U}_{d}^{\top}\boldsymbol{U}_{d}=\boldsymbol{I}_{I_{d}},P_{d}<I_{d}\}_{d=1}^D$
such that the projected low-dimensional tensor captures most of the
variation in the original tensor. Mathematically, this can be formalized
into the following optimization problem:

\begin{equation}
\{\boldsymbol{U}_{1},\ldots,\boldsymbol{U}_{D}\}=\argmax_{\boldsymbol{U}_{1},\ldots,\boldsymbol{U}_{D}}\sum_{i=1}^{N}\left\Vert (\mathcal{S}_{i}-\bar{\mathcal{S}})\times_{1}\boldsymbol{U}_{1}^{\top}\times_{2}\boldsymbol{U}_{2}^{\top}\times\cdots\times_{D}\boldsymbol{U}_{D}^{\top}\right\Vert _{F}^{2}\label{eq:mpca}
\end{equation}

\noindent where $\bar{\mathcal{S}}=\frac{1}{N}\sum_{i=1}^{N}\mathcal{S}_{i}$
is the mean tensor. This optimization problem can be solved iteratively
using the algorithm given in Appendix \ref{app:Optimization-mpca}.
Additional details regarding the algorithm and the methods used to
determine the dimensionality of the tensor subspace, $\{P_{d}\}_{d=1}^{D}$,
can be found in \citet{Lu2008}. This approach helps to reduce the
number of parameters to be estimated from $2+\prod_{d=1}^{D}I_{d}$
in Equation (\ref{eq:treg}) to $2+\prod_{d=1}^{D}P_{d}$ in Equation
(\ref{eq:treg2}) where $2+\prod_{d=1}^{D}P_{d}<<2+\prod_{d=1}^{D}I_{d}$.

However, often, the number of reduced parameters (i.e., $2+\prod_{d=1}^{D}P_{d}$)
is still so large that requires further dimension reduction. For example,
for a $40\times50\times100$ tensor, if $P_{1}=P_{2}=P_{3}=10$, the
number of parameters is reduced from 200,002 to 1,002. To further
reduce the number of parameters so that they can be estimated by using
a limited training sample, we utilize two well-known tensor decomposition
techniques namely, CP and Tucker decompositions. We briefly review
these decompositions in Sections \ref{sec:cp} and \ref{sec:tucker},
and discuss how they are incorporated into our prognostic framework.

\subsection{Dimension Reduction via CP Decomposition}

\label{sec:cp}

In CP decomposition, the coefficient tensor $\tilde{\mathcal{B}}$
in Equation (\ref{eq:treg2}) is decomposed into a sum product of
a set of rank one vectors. Given the rank of $\tilde{\mathcal{B}}$,
which we denote by $R$, we have the following decomposition,

\begin{equation}
\tilde{\mathcal{B}}=\sum_{r=1}^{R}\tilde{\boldsymbol{\beta}}_{1}^{(r)}\circ\cdots\circ\tilde{\boldsymbol{\beta}}_{D}^{(r)},\label{eq:cp}
\end{equation}

\noindent where $\tilde{\boldsymbol{\beta}}_{d}^{(r)}=\left[\tilde{\beta}_{d,1}^{(r)},\ldots,\tilde{\beta}_{d,P_{d}}^{(r)}\right]^{\top}\in\mathbb{R}^{P_{d}}$,
and ``$\circ$" denotes the outer product operator. It can be easily
shown that ${vec}(\tilde{\mathcal{B}})=(\tilde{\boldsymbol{B}}_{D}\odot\cdots\odot\tilde{\boldsymbol{B}}_{1})\boldsymbol{1}_{R}$,
where $\tilde{\boldsymbol{B}}_{d}=\left[\tilde{\boldsymbol{\beta}}_{d}^{(1)},\ldots,\tilde{\boldsymbol{\beta}}_{d}^{(R)}\right]\in\mathbb{R}^{P_{d}\times R}$
for $d=1,\ldots,D$ and $\boldsymbol{1}_{R}\in\mathbb{R}^{R}$ is
an $R$ dimensional vector of ones. Thus, Equation (\ref{eq:treg2})
can be re-expressed as follows:

\begin{equation}
\begin{split}y_{i} & =\alpha+\left\langle {vec}(\tilde{\mathcal{B}}),{vec}(\tilde{\mathcal{S}}_{i})\right\rangle +\sigma\epsilon_{i}\\
 & =\alpha+\left\langle (\tilde{\boldsymbol{B}}_{D}\odot\cdots\odot\tilde{\boldsymbol{B}}_{1})\boldsymbol{1}_{R},{vec}(\tilde{\mathcal{S}}_{i})\right\rangle +\sigma\epsilon_{i}\\
\\
\end{split}
\label{eq:treg3}
\end{equation}

The number of parameters in Equation (\ref{eq:treg3}) is $2+\sum_{d=1}^{D}P_{d}\times R$,
which is significantly smaller than $2+\prod_{d=1}^{D}P_{d}$ from
(\ref{eq:treg2}). In our 3-order tensor example, if $P_{1}=P_{2}=P_{3}=10$
and the rank $R=2$, the number of parameters decreases from $1,002$
to $62=2+10\times2+10\times2+10\times2$.

\subsubsection{Parameter Estimation for CP Decomposition}

\label{subsec:cpest} To estimate the parameters of Equation (\ref{eq:treg3})
using MLE, we maximize the corresponding penalized log-likelihood
function:

\begin{align}
 & \argmax_{\boldsymbol{\theta}}\left\{ \ell(\boldsymbol{\theta})-\sum_{d=1}^{D}r\left(\tilde{\boldsymbol{B}}_{d}\right)\right\} \label{eq:lmle-1}\nonumber\\
  =&\argmax_{\boldsymbol{\theta}}\left\{ -N\log\sigma+\sum_{i=1}^{N}\log f\left(\frac{y_{i}-\alpha-\left\langle (\tilde{\boldsymbol{B}}_{D}\odot\cdots\odot\tilde{\boldsymbol{B}}_{1})\boldsymbol{1}_{R},{vec}(\tilde{\mathcal{S}}_{i})\right\rangle }{\sigma}\right)\right.\nonumber\\
&\left.-\sum_{d=1}^{D}r\left(\tilde{\boldsymbol{B}}_{d}\right)\right\}
\end{align}

\noindent where $\boldsymbol{\theta}=(\alpha,\sigma,\tilde{\boldsymbol{B}}_{1}\ldots,\tilde{\boldsymbol{B}}_{D})$
and $r(\tilde{\boldsymbol{B}}_{d})=\lambda_{d}\sum_{r=1}^{R}\sum_{j=1}^{P_{d}}\|\tilde{\beta}_{d,j}^{(r)}\|_{1}$.
The $\ell_{1}$-norm penalty term encourages the sparsity of $\tilde{\mathcal{B}}$,
which helps avoid over-fitting.

The block relaxation method proposed by \citep{De1994,Lange2010}
is used to maximize expression (\ref{eq:lmle-1}). Specifically, we
iteratively update a block of parameters, say $(\tilde{\boldsymbol{B}}_{d},\sigma,\alpha)$,
while keeping other components $\{\tilde{\boldsymbol{B}}_{1},\ldots,\tilde{\boldsymbol{B}}_{d-1},\tilde{\boldsymbol{B}}_{d+1},\ldots,\tilde{\boldsymbol{B}}_{D}\}$
fixed. In each update, the optimization criterion is reduced to $\argmax_{\boldsymbol{\tilde{B}}_{d},\sigma,\alpha}\left\{ \ell(\boldsymbol{\theta})-r(\tilde{\boldsymbol{B}}_{d})\right\} $.

Next, we show in Proposition \ref{prop:cp} that the optimization
problem for each block $\tilde{\boldsymbol{B}}_{d}$ is equivalent
to optimizing the penalized log-likelihood function for $y_{i}=\alpha+\langle\tilde{\boldsymbol{B}}_{d},\boldsymbol{X}_{d,i}\rangle+\sigma\epsilon_{i}$,
where $\tilde{\boldsymbol{B}}_{d}$ is the parameter matrix and $\boldsymbol{X}_{d,i}$
is the predictor matrix defined by $\boldsymbol{X}_{d,i}=\tilde{\boldsymbol{S}}_{i(d)}(\tilde{\boldsymbol{B}}_{D}\odot\cdots\odot\tilde{\boldsymbol{B}}_{d+1}\odot\tilde{\boldsymbol{B}}_{d-1}\odot\cdots\odot\tilde{\boldsymbol{B}}_{1})$,
and where $\tilde{\boldsymbol{S}}_{i(d)}$ is the mode-\textit{d}
matricization of $\tilde{\mathcal{S}}_{i}$ (defined in the Preliminaries
Section). 
\begin{prop}
\label{prop:cp}Consider the optimization problem in (\ref{eq:lmle-1}),
given $(\tilde{\boldsymbol{B}}_{1},\ldots,\tilde{\boldsymbol{B}}_{d-1},\tilde{\boldsymbol{B}}_{d+1},\ldots,\tilde{\boldsymbol{B}}_{D})$,
the optimization problem can be reduced to 
\end{prop}
\begin{equation}
\argmax_{\boldsymbol{\tilde{B}}_{d},\sigma,\alpha}\left\{ -N\ln\sigma+\sum_{i=1}^{N}\ln f\left(\frac{y_{i}-\alpha-\left\langle \tilde{\boldsymbol{B}}_{d},\boldsymbol{X}_{d,i}\right\rangle }{\sigma}\right)-r(\tilde{\boldsymbol{B}}_{d})\right\} .\label{eq:bd3-1}
\end{equation}

The proof of Proposition \ref{prop:cp} is provided in Appendix \ref{app:Proof-for-Propositioncp}.
As pointed out by \citet{Stadler2010}, the estimates of $\alpha,\boldsymbol{\tilde{B}}_{d},\sigma$
in optimizing problem (\ref{eq:bd3-1}) are not invariant under scaling
of the response. To be specific, consider the transformation $y_{i}'=by_{i},\alpha'=b\alpha,\tilde{\boldsymbol{B}}_{d}'=b\tilde{\boldsymbol{B}_{d}},\sigma'=b\sigma$
where $b>0$. Clearly, this transformation does not affect the regression
model $y_{i}=\alpha+\langle\tilde{\boldsymbol{B}}_{d},\boldsymbol{X}_{d,i}\rangle+\sigma\epsilon_{i}$.
Therefore, invariant estimates based on the transformed data $(y_{i}',\boldsymbol{X}_{d,i})$,
should satisfy $\hat{\alpha}'=b\hat{\alpha},\hat{\tilde{\boldsymbol{B}}}_{d}'=b\hat{\tilde{\boldsymbol{B}}}_{d},\hat{\sigma}'=b\hat{\sigma}$,
where $\hat{\alpha},\hat{\tilde{\boldsymbol{B}}}_{d},\hat{\sigma}$
are estimates based on original data $(y_{i},\boldsymbol{X}_{d,i})$.
However, this does not hold for the estimates obtained by optimizing
(\ref{eq:bd3-1}). To address this issue, expression (\ref{eq:bd3-1})
is modified by dividing the penalty term by the scale parameter $\sigma$:

\begin{equation}
\argmax_{\boldsymbol{\tilde{B}}_{d},\sigma,\alpha}\left\{ -N\ln\sigma+\sum_{i=1}^{N}\ln f\left(\frac{y_{i}-\alpha-\left\langle \tilde{\boldsymbol{B}}_{d},\boldsymbol{X}_{d,i}\right\rangle }{\sigma}\right)-r(\frac{\tilde{\boldsymbol{B}}_{d}}{\sigma})\right\} .\label{eq:bd4}
\end{equation}

We can show that the resulting estimates possess the invariant property
(see Appendix \ref{app:invariant}). Note that in the modified problem,
the penalty term penalizes the $\ell_{1}$-norm of the coefficients
and the scale parameter $\sigma$ simultaneously, which has some close
relations to the Bayesian Lasso \citep{Park2008,Stadler2010}. The
log-likelihood function in (\ref{eq:bd4}) is not concave which causes
computational problems. We use the following re-parameterization to
transform the optimization function to a concave function: $\alpha_{0}=\alpha/\sigma,\boldsymbol{A}_{d}=\tilde{\boldsymbol{B}}_{d}/\sigma,\rho=1/\sigma$.
Consequently, the optimization problem can be rewritten as:

\begin{equation}
\argmax_{\boldsymbol{A}_{d},\rho,\alpha_{0}}\left\{ N\ln\rho+\sum_{i=1}^{N}\ln f\left(\rho y_{i}-\alpha_{0}-\left\langle \boldsymbol{A}_{d},\boldsymbol{X}_{d,i}\right\rangle \right)-r(\boldsymbol{A}_{d})\right\} .\label{eq:bd5}
\end{equation}

The optimization problem in (\ref{eq:bd5}) is concave if function
$f$ is log-concave, which is the case for most LLS distributions
including \textit{normal}, \textit{logistic}, \textit{SEV}, \textit{generalized
log-gamma}, \textit{log-inverse Gaussian} \citep{Doray94}. \textit{Lognormal},
\textit{log-logistic} and \textit{Weibull} distributions whose density
function is not log-concave can easily be transformed to \textit{normal},
\textit{logistic} and \textit{SEV} distributions, respectively, by
taking the logarithm of the TTF. Various optimization algorithms such
as coordinate descent \citep{Friedman2007} and gradient descent \citep{Tseng2001} can be used for solving (\ref{eq:bd5}). Algorithm 1 shows
the steps of the block relaxation method for optimizing (\ref{eq:bd5})
and finding the ML estimates of the parameters. 

\begin{table}[htb!]
\small
\centering %
\begin{tabular}{llll}
\hline 
\textbf{Algorithm 1}: Block relaxation algorithm for solving problem
(\ref{eq:lmle-1}).  &  &  & \tabularnewline
\hline 
\textbf{Input:} $\{\tilde{\mathcal{S}_{i}},y_{i}\}_{i=1}^{N}$ and
rank $R$  &  &  & \tabularnewline
\textbf{Initialization:} Matrices $\tilde{\boldsymbol{B}}_{2}^{(0)},\tilde{\boldsymbol{B}}_{3}^{(0)},\ldots,\tilde{\boldsymbol{B}}_{D}^{(0)}$
are initialized randomly.  &  &  & \tabularnewline
\textbf{while} convergence criterion not met \textbf{do}  &  &  & \tabularnewline
\;\;\textbf{for} $d=1,\ldots,D$ \textbf{do}  &  &  & \tabularnewline
\;\;\;\;\;\;$\boldsymbol{X}_{d,i}^{(k+1)}=\tilde{\boldsymbol{S}}_{i(d)}(\tilde{\boldsymbol{B}}_{D}^{(k)}\odot\cdots\odot\tilde{\boldsymbol{B}}_{d+1}^{(k)}\odot\tilde{\boldsymbol{B}}_{d-1}^{(k+1)}\odot\cdots\odot\tilde{\boldsymbol{B}}_{1}^{(k+1)})$  &  &  & \tabularnewline
\;\;\;\; $\begin{aligned}\boldsymbol{A}_{d}^{(k+1)},\rho^{(k+1)},\alpha_{0}^{(k+1)}=\argmax_{\boldsymbol{A}_{d},\rho,\alpha_{0}}\{N\ln\rho+\sum_{i=1}^{N}\ln f(\rho y_{i}-\alpha_{0}-\langle\boldsymbol{A}_{d},\boldsymbol{X}_{d,i}^{(k+1)}\rangle)-r(\boldsymbol{A}_{d})\}\end{aligned}
$  &  &  & \tabularnewline
\;\;\;\; $\tilde{\boldsymbol{B}}_{d}^{(k+1)}=\boldsymbol{A}_{d}^{(k+1)}/\rho^{(k+1)}$  &  &  & \tabularnewline
\;\;\textbf{end for}  &  &  & \tabularnewline
\;\;Let $k=k+1$  &  &  & \tabularnewline
\textbf{end while}  &  &  & \tabularnewline
\textbf{Output}:$\alpha=\alpha_{0}/\rho,\sigma=1/\rho,\{\tilde{\boldsymbol{B}}_{d}\}_{d=1}^{D}$  &  &  & \tabularnewline
\hline 
\end{tabular}\label{table:algorithm} 
\end{table}

The convergence criterion is defined by $\ell(\tilde{\boldsymbol{\theta}}^{(k+1)})-\ell(\tilde{\boldsymbol{\theta}}^{(k)})<\epsilon$,
in which $\ell(\tilde{\boldsymbol{\theta}}^{(k)})$, is defined as
follows:

\begin{equation}
\begin{aligned}
\ell(\tilde{\boldsymbol{\theta}}^{(k)})=-N\log\sigma^{(k)}+\sum_{i=1}^{N}\log f\left(\frac{y_{i}-\alpha^{(k)}-\left\langle (\tilde{\boldsymbol{B}}_{D}^{(k)}\odot\cdots\odot\tilde{\boldsymbol{B}}_{1}^{(k)})\boldsymbol{1}_{R},{vec}(\tilde{\mathcal{S}}_{i})\right\rangle }{\sigma^{(k)}}\right)\\-\sum_{d=1}^{D}r\left(\frac{\tilde{\boldsymbol{B}}_{d}^{(k)}}{\sigma^{(k)}}\right)\label{eq:ov}
\end{aligned}
\end{equation}

\noindent where $\tilde{\boldsymbol{\theta}}^{(k)}=(\alpha^{(k)},\sigma^{(k)},\tilde{\boldsymbol{B}}_{1}^{(k)},\ldots,\tilde{\boldsymbol{B}}_{D}^{(k)})$.

It can be shown that Algorithm 1 exhibits the global convergence property
(see Proposition 1 in \citet{Zhou13}). In other words, it will converge
to a stationary point for any initial point. Since a stationary point
is only guaranteed to be a local maximum or saddle point, the algorithm
is run several times with different initializations while recording
the best results.

Algorithm 1 requires the rank of $\tilde{\mathcal{B}}$ to be known
a priori for CP decomposition. In this paper, the Bayesian information
criterion (BIC) is used to determine the appropriate rank. The BIC
is defined as $-2\ell(\tilde{\boldsymbol{\theta}})+P\ln(N)$, where
$\ell$ is the log-likelihood value defined in Equation (\ref{eq:ov}),
$N$ is the sample size (number of systems) and $P$ is the number
of effective parameters. For $D=2$, we set $P=R(P_{1}+P_{2})-R^{2}$,
where $-R^{2}$ is used for adjustment of the nonsingular transformation
indeterminacy for model identifiability; for $D>2$, we set $P=R(\sum_{d=1}^{D}P_{d}-D+1)$,
where $R(-D+1)$ is used for the scaling indeterminacy in the CP decomposition
\citep{Li2013}.

\subsection{Dimension Reduction via Tucker decomposition}

\label{sec:tucker}

Tucker decomposition is the second tensor decomposition approach used
in this paper. It is used to reduce the dimensionality of $\tilde{\mathcal{B}}$
as a product of a low-dimensional core tensor and a set of factor
matrices as follows:

\begin{equation}
\tilde{\mathcal{B}}=\mathcal{\tilde{G}}\times_{1}\tilde{\boldsymbol{B}}_{1}\times_{2}\tilde{\boldsymbol{B}}_{2}\times_{3}\cdots\times_{D}\tilde{\boldsymbol{B}}_{D}=\sum_{r_{1}=1}^{R_{1}}\cdots\sum_{r_{D}=1}^{R_{D}}\tilde{g}_{r_{1},\ldots,r_{D}}\tilde{\boldsymbol{\beta}}_{1}^{(r_{1})}\circ\cdots\circ\tilde{\boldsymbol{\beta}}_{D}^{(r_{D})},\label{eq:tucker}
\end{equation}

\noindent where $\mathcal{\tilde{G}}\in\mathbb{R}^{R_{1}\times R_{2}\times\cdots\times R_{D}}$
is the core tensor with element $(\tilde{\mathcal{G}})_{r_{1},\ldots,r_{D}}=\tilde{g}_{r_{1},\ldots,r_{D}}$,
$\tilde{\boldsymbol{B}}_{d}=\left[\tilde{\boldsymbol{\beta}}_{d}^{(1)},\ldots,\tilde{\boldsymbol{\beta}}_{d}^{(R_{d})}\right]\in\mathbb{R}^{P_{d}\times R_{d}}$
for $d=1,\ldots,D$ is the factor matrix, ``$\times_{d}$'' is the
mode-$d$ product operator, and ``$\circ$'' is the outer product
operator. Using this decomposition, Equation (\ref{eq:treg2}) can
be re-expressed as follows:

\begin{equation}
\begin{split}y_{i} & =\alpha+\langle\tilde{\mathcal{B}},\tilde{\mathcal{S}}_{i}\rangle+\sigma\epsilon_{i}\\
 & =\alpha+\langle\mathcal{\tilde{G}}\times_{1}\tilde{\boldsymbol{B}}_{1}\times_{2}\tilde{\boldsymbol{B}}_{2}\times_{3}\cdots\times_{D}\tilde{\boldsymbol{B}}_{D},\tilde{\mathcal{S}}_{i}\rangle+\sigma\epsilon_{i}
\end{split}
\label{eq:tuckerreg1}
\end{equation}

\subsubsection{Parameter Estimation for Tucker Decomposition}

The following penalized log-likelihood function is used to compute
the MLE estimates of the parameters in expression (\ref{eq:tuckerreg1}).

\begin{align}
 & \argmax_{\boldsymbol{\theta}}\left\{ \ell(\boldsymbol{\theta})-r(\mathcal{\mathcal{\tilde{G}}})-\sum_{d=1}^{D}r\left(\tilde{\boldsymbol{B}}_{d}\right)\right\} \nonumber \\
= & \argmax_{\boldsymbol{\theta}}\left\{ -N\ln\sigma+\sum_{i=1}^{N}\ln f\left(\frac{y_{i}-\alpha-\left\langle \mathcal{\tilde{G}}\times_{1}\tilde{\boldsymbol{B}}_{1}\times_{2}\tilde{\boldsymbol{B}}_{2}\times_{3}\cdots\times_{D}\tilde{\boldsymbol{B}}_{D},\tilde{\mathcal{S}}_{i}\right\rangle }{\sigma}\right)\right.\nonumber \\
 & \left.-r(\mathcal{\mathcal{\tilde{G}}})-\sum_{d=1}^{D}r\left(\tilde{\boldsymbol{B}}_{d}\right)\right\} ,\label{eq:tuckerlmle2}
\end{align}

\noindent where $\boldsymbol{\theta}=(\alpha,\sigma,\mathcal{\mathcal{\tilde{G}}},\tilde{\boldsymbol{B}}_{1},\ldots,\tilde{\boldsymbol{B}}_{D})$, $r(\mathcal{\mathcal{\mathcal{\tilde{G}}}})=\lambda\sum_{r_{1}=1}^{R_{1}}\cdots\sum_{r_{D}=1}^{R_{D}}\|\tilde{g}_{r_{1},\ldots,r_{D}}\|_{1}$
and \;$r(\tilde{\boldsymbol{B}}_{d})=\lambda_{d}\sum_{r_{d}=1}^{R_{d}}\sum_{j=1}^{P_{d}}\|\tilde{\beta}_{d,j}^{(r_{d})}\|_{1}$.

Similar to the CP decomposition model, the block relaxation method
is used to solve expression (\ref{eq:tuckerlmle2}). To update the
core tensor $\mathcal{\mathcal{\tilde{G}}}$ given all the factor
matrices, the optimization criterion is reduced to $\argmax_{\mathcal{\tilde{G}}}\left\{ \ell(\boldsymbol{\theta})-r(\mathcal{\mathcal{\tilde{G}}})\right\} .$
Proposition 3 shows that this optimization problem is equivalent to
optimizing the penalized log-likelihood function of $y_{i}=\alpha+\langle vec(\mathcal{\mathcal{\mathcal{\tilde{G}}}}),\boldsymbol{x}_{i}\rangle+\sigma\epsilon_{i}$,
where $vec(\mathcal{\mathcal{\mathcal{\tilde{G}}}})$ is the parameter
vector and $\boldsymbol{x}_{i}$ is the predictor vector defined by
$\boldsymbol{x}_{i}=(\tilde{\boldsymbol{B}}_{D}\otimes\cdots\otimes\tilde{\boldsymbol{B}}_{1})^{\top}vec(\tilde{\mathcal{S}}_{i})$. 
\begin{prop}
\label{prop:tucker1}Consider the optimization problem in (\ref{eq:tuckerlmle2}),
given $\{\tilde{\boldsymbol{B}}_{1},\ldots,\tilde{\boldsymbol{B}}_{D}\}$,
the optimization problem is reduced to 
\begin{equation}
\argmax_{\mathcal{\mathcal{\mathcal{\mathcal{\tilde{G}}}}}}\left\{ -N\ln\sigma+\sum_{i=1}^{N}\ln f\left(\frac{y_{i}-\alpha-\left\langle vec(\mathcal{\mathcal{\mathcal{\mathcal{\tilde{G}}}}}),(\tilde{\boldsymbol{B}}_{D}\otimes\cdots\otimes\tilde{\boldsymbol{B}}_{1})^{\top}vec(\tilde{\mathcal{S}}_{i})\right\rangle }{\sigma}\right)-r(\mathcal{\mathcal{\mathcal{\mathcal{\tilde{G}}}}})\right\} ,\label{eq:tuckercore}
\end{equation}
\end{prop}
The proof of Proposition 3 is given in Appendix \ref{app:Proof-for-Propositiontucker}.
To guarantee the invariance property of the estimates and concavity
of the optimization function, we apply the following reparameterization:
$\rho=1/\sigma,\alpha_{0}=\alpha/\sigma,\mathcal{C}=\mathcal{\tilde{\mathcal{G}}}/\sigma$,$r(\mathcal{C})=\lambda\sum_{r_{1}=1}^{R_{1}}\cdots\sum_{r_{D}=1}^{R_{D}}\frac{\|\tilde{g}_{r_{1},\ldots,r_{D}}\|_{1}}{\sigma}$.
This enables us to re-express criterion (\ref{eq:tuckercore}) as
follows:

\begin{equation}
\argmax_{\mathcal{C},\rho,\alpha_{0}}\left\{ N\ln\rho+\sum_{i=1}^{N}\ln f\left(\rho y_{i}-\alpha_{0}-\left\langle vec(\mathcal{C}),(\tilde{\boldsymbol{B}}_{D}\otimes\cdots\otimes\tilde{\boldsymbol{B}}_{1})^{\top}vec(\tilde{\mathcal{S}}_{i})\right\rangle \right)-r(\mathcal{C})\right\} .\label{eq:g3}
\end{equation}

To update the factor matrix $\tilde{\boldsymbol{B}}_{d}$ for $d=1,\ldots,D$,
we fix the core tensor $\mathcal{\mathcal{\tilde{\mathcal{G}}}}$
and the rest of the factor matrices $\{\tilde{\boldsymbol{B}}_{1},\ldots,\tilde{\boldsymbol{B}}_{d-1},\tilde{\boldsymbol{B}}_{d+1}\ldots,\tilde{\boldsymbol{B}}_{D}\}$,
and maximize the following criterion $\argmax_{\tilde{\boldsymbol{B}}_{d}}\left\{ \ell(\boldsymbol{\theta})-r(\tilde{\boldsymbol{B}}_{d})\right\} $.
Proposition \ref{prop:tucker2} shows that this optimization problem
is equivalent to optimizing the log-likelihood function of $y_{i}=\alpha+\langle\tilde{\boldsymbol{B}}_{d},\boldsymbol{X}_{d,i}\rangle+\sigma\epsilon_{i},$where
$\tilde{\boldsymbol{B}}_{d}$ is the parameter matrix and $\boldsymbol{X}_{d,i}$
is the predictor matrix defined by $\boldsymbol{X}_{d,i}=\tilde{\boldsymbol{S}}_{i(d)}(\tilde{\boldsymbol{B}}_{D}\otimes\cdots\otimes\tilde{\boldsymbol{B}}_{d+1}\otimes\tilde{\boldsymbol{B}}_{d-1}\otimes\cdots\otimes\tilde{\boldsymbol{B}}_{1})\tilde{\boldsymbol{G}}_{(d)}^{\top}$,
where $\tilde{\boldsymbol{S}}_{i(d)}$ and $\tilde{\boldsymbol{G}}_{(d)}$
are the mode-\emph{d} matricization of $\tilde{\mathcal{S}_{i}}$
and $\tilde{\mathcal{G}}$, respectively. 

\begin{prop}
\label{prop:tucker2} Consider the problem in (\ref{eq:tuckerlmle2}),
given $\mathcal{\tilde{G}}$ and $\{\tilde{\boldsymbol{B}}_{1},\ldots,\tilde{\boldsymbol{B}}_{d-1},\tilde{\boldsymbol{B}}_{d+1},\ldots,\tilde{\boldsymbol{B}}_{D}\}$,
the optimization problem is reduced to 
\begin{equation}
\argmax_{\tilde{\boldsymbol{B}}_{d}}\left\{ -N\ln\sigma+\sum_{i=1}^{N}\ln f\left(\frac{y_{i}-\alpha-\left\langle \tilde{\boldsymbol{B}}_{d},\boldsymbol{X}_{d,i}\right\rangle }{\sigma}\right)-r(\tilde{\boldsymbol{B}}_{d})\right\}. \label{eq:tuckerbasis}
\end{equation}
\end{prop}

The proof of Proposition \ref{prop:tucker2} is provided in Appendix
\ref{app:Proof-for-Propositiontucker2}. Similar to expression (\ref{eq:bd4}),
we use penalty term $r(\frac{\tilde{\boldsymbol{B}}_{d}}{\sigma})$
and let $\rho=1/\sigma,\alpha_{0}=\alpha/\sigma,\boldsymbol{A}_{d}=\tilde{\boldsymbol{B}}_{d}/\sigma$.
Consequently, we obtain the following optimization subproblem for
parameter estimation:

\begin{equation}
\argmax_{\boldsymbol{A}_{d}}\left\{ N\ln\rho+\sum_{i=1}^{N}\ln f\left(\rho y_{i}-\alpha_{0}-\left\langle \boldsymbol{A}_{d},\boldsymbol{X}_{d,i}\right\rangle \right)-r(\boldsymbol{A}_{d})\right\} .\label{eq:b3}
\end{equation}

\begin{table}[htb!]
\small
\centering %
\begin{tabular}{l}
\hline 
\textbf{Algorithm 2}: Block relaxation algorithm for solving problem
(\ref{eq:tuckerreg1}). \tabularnewline
\hline 
\textbf{Input:} $\{\tilde{\mathcal{S}_{i}},y_{i}\}_{i=1}^{N}$ and
rank $\{R_{d}\}_{d=1}^{D}$\tabularnewline
\textbf{Initialization:} Core tensor $\mathcal{G}^{(0)}$ and matrices
$\tilde{\boldsymbol{B}}_{2}^{(0)},\tilde{\boldsymbol{B}}_{3}^{(0)},\ldots,\tilde{\boldsymbol{B}}_{D}^{(0)}$
are initialized randomly.\tabularnewline
\textbf{while} convergence criterion not met \textbf{do} \tabularnewline
\;\;\textbf{for} $d=1,\ldots,D$ \textbf{do}\tabularnewline
\;\;\;\;$\boldsymbol{X}_{d,i}^{(k+1)}=\tilde{\boldsymbol{S}}_{i(d)}(\tilde{\boldsymbol{B}}_{D}^{(k)}\otimes\cdots\otimes\tilde{\boldsymbol{B}}_{d+1}^{(k)}\otimes\tilde{\boldsymbol{B}}_{d-1}^{(k+1)}\otimes\cdots\otimes\tilde{\boldsymbol{B}}_{1}^{(k+1)})\{\boldsymbol{G}_{(d)}^{(k)}\}^{\top}$\tabularnewline
\;\; $\begin{aligned}\boldsymbol{A}_{d}^{(k+1)},\rho^{(k+1)},\alpha_{0}^{(k+1)}=\argmax_{\boldsymbol{A}_{d},\rho,\alpha_{0}}\{N\ln\rho+\sum_{i=1}^{N}\ln f(\rho y_{i}-\alpha_{0}-\langle\boldsymbol{A}_{d},\boldsymbol{X}_{d,i}^{(k+1)}\rangle)-r(\boldsymbol{A}_{d})\}\end{aligned}
$\tabularnewline
\;\; $\tilde{\boldsymbol{B}}_{d}^{(k+1)}=\boldsymbol{A}_{d}^{(k+1)}/\rho^{(k+1)}$\tabularnewline
\;\;\textbf{end for}\tabularnewline
\;\;$\begin{aligned}\mathcal{C}^{(k+1)},\rho^{(k+1)},\alpha_{0}^{(k+1)}=\argmax_{\mathcal{C},\rho,\alpha_{0}}\{N\ln\rho+\sum_{i=1}^{N}\ln f(\rho y_{i}-\alpha_{0}-\langle vec(\mathcal{C}),(\tilde{\boldsymbol{B}}_{D}^{(k+1)}\otimes\cdots\\
\otimes\tilde{\boldsymbol{B}}_{1}^{(k+1)})^{\top}vec(\tilde{\mathcal{S}}_{i})\rangle)-r(\mathcal{C})\}
\end{aligned}
$\tabularnewline
\;\;$\mathcal{G}^{(k+1)}=\mathcal{C}^{(k+1)}/\rho^{(k+1)}$\tabularnewline
\;\;Let $k=k+1$\tabularnewline
\textbf{end while}\tabularnewline
\textbf{Output}: $\alpha=\alpha_{0}/\rho,\sigma=1/\rho,\mathcal{G},\{\tilde{\boldsymbol{B}}_{d}\}_{d=1}^{D}$\tabularnewline
\hline 
\end{tabular}\label{table:algorithm2} 
\end{table}

The pseudocode for the block relaxation algorithm is summarized in
Algorithm 2. The convergence criterion is defined by $\ell(\tilde{\boldsymbol{\theta}}^{(k+1)})-\ell(\tilde{\boldsymbol{\theta}}^{(k)})<\epsilon$,
where $\ell(\tilde{\boldsymbol{\theta}}^{(k)})$, is given by

\begin{equation}
\begin{aligned}\ell(\tilde{\boldsymbol{\theta}}^{(k)})=-N\ln\sigma^{(k)}+\sum_{i=1}^{N}\ln f\left(\frac{y_{i}-\alpha^{(k)}-\left\langle \mathcal{\tilde{G}}^{(k)}\times_{1}\tilde{\boldsymbol{B}}_{1}^{(k)}\times_{2}\tilde{\boldsymbol{B}}_{2}^{(k)}\times_{3}\cdots\times_{D}\tilde{\boldsymbol{B}}_{D}^{(k)},\tilde{\mathcal{S}}_{i}\right\rangle }{\sigma}\right)\\
-r(\mathcal{\mathcal{\tilde{G}}}^{(k)})-\sum_{d=1}^{D}r\left(\tilde{\boldsymbol{B}}_{d}^{(k)}\right),
\end{aligned}
\label{eq:ov2}
\end{equation}

\noindent where $\tilde{\boldsymbol{\theta}}^{(k)}=(\alpha^{(k)},\rho^{(k)},\mathcal{G}^{(k)},\tilde{\boldsymbol{B}}_{1}^{(k)},\ldots,\tilde{\boldsymbol{B}}_{D}^{(k)})$.

The set of ranks (i.e., $R_{1},R_{2},\cdots,R_{D}$) used in the Tucker
decomposition is an input to Algorithm 2. BIC is also used here to
determine the appropriate rank, where $\ell$ is the log-likelihood
value defined in Equation (\ref{eq:ov2}), $N$ is the sample size
(number of systems) and $P=\sum_{d=1}^{D}P_{d}R_{d}+\prod_{d=1}^{D}R_{d}-\sum_{d=1}^{D}R_{d}^{2}$
is the number of effective parameters. Here the term $-\sum_{d=1}^{D}R_{d}^{2}$
is used to adjust for the non-singular transformation indeterminacy
in the Tucker decomposition \citep{Li2013}.

Using BIC for rank selection in the Tucker-based tensor regression
model can be computationally prohibitive. For example, for a 3-order
tensor, there are totally $27=3^{3}$ rank candidates when the maximum
rank in each dimensionality is 3. Increasing the maximum rank to 4
and 5, the number of rank candidates is increased to $64=4^{3}$ and
$125=5^{3}$, respectively. To address this challenge, we propose
a computationally efficient heuristic method that automatically selects
an appropriate rank. First, an initial coefficient tensor is estimated
by regressing each pixel against the TTF. Next, high-order singular
value decomposition (HOSVD) \citep{De Lathauwer2000} is applied to
the estimated tensor. HOSVD works by applying regular SVD to matricizations
of the initial tensor on each mode. The rank of each mode can be selected
by using fraction-of-variance explained (FVE) \citep{Fang2015} and
the resulting eigenvector matrix is the factor matrix for that mode.
Given the initial tensor and its estimated factor matrices, we can
estimate the core tensor. The core tensor and factor matrices estimated
by HOSVD are used for initialization in Algorithm 2. As pointed out
by  various studies in the literature, HOSVD often performs reasonably
well as an initialization method for iterative tensor estimation algorithms
\citep{Kolda2009,Lu2008}.

\section{RUL prediction and realtime updating}

\label{sec:updating}

The goal of this paper is to predict and update the RUL of partially
degraded systems using in-situ degradation image streams. To achieve
this, we utilize the LLS regression modeling framework discussed in
Section 3, and update the trained model based on data streams observed
from fielded systems. The LLS regression model requires that the degradation
image streams of the training systems and the fielded system being
monitored to have the same dimensionality. In other words, both should
have the same number of degradation images or profile length. In reality,
this attribute is difficult to maintain for two reasons; (1) different
systems have different failure times, and (2) an equipment is typically
shutdown after failure and no further observations can be made beyond
the failure time. Assuming the sampling (observation) time intervals
are the same for all systems, a system with a longer failure time
has more degradation data than a system with a short failure time.

\begin{figure}
\begin{centering}
\includegraphics[scale=0.5]{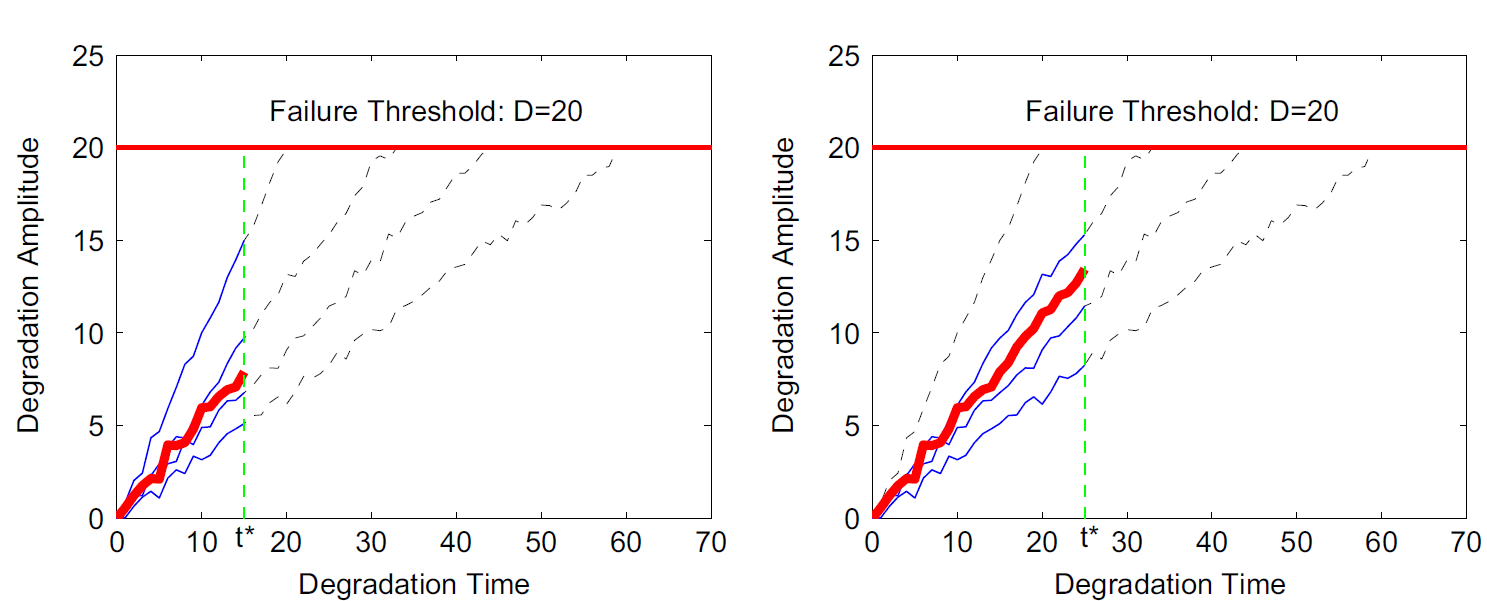} 
\par\end{centering}
\caption{Method of updating the model as time advances \citep{Fang2015}}
\label{fig:updating} 
\end{figure}

To address this challenge, we adopt the time-varying regression framework
used in \citet{Fang2015}. The idea of the time-varying regression
is that systems whose TTF are shorter than the current observation
time (of the fielded system) are excluded from the training dataset.
Next, the degradation data of the chosen systems are truncated at
the current observation time as illustrated in Figure \ref{fig:updating}.
By doing this, we ensure that the truncated degradation tensors of
the chosen systems and the real-time observed degradation tensors
of the fielded system possess the same dimensionality.

We summarize the process of predicting and updating the RUL of a fielded
system as follows: 
\begin{enumerate}[(i)]
\item At each sampling time $t_{n}$, a new degradation image is observed
from a fielded system. Systems whose TTF are longer than $t_{n}$
are chosen from the training dataset. 
\item The image streams of the chosen systems are then truncated at time
$t_{n}$ by keeping only the images observed at times $\{t_{1},t_{2},\ldots,t_{n}\}$.
The truncated image streams constitutes a new ``training dataset'',
hereafter referred to as \textit{truncated training dataset}.
\item A dimensionality reduction technique, such as MPCA, is applied to
the \textit{truncated training dataset} to obtain a low-dimensional
tensor subspace of the \textit{truncated training dataset}. Tensors
in the \textit{truncated training dataset} are then projected to the
tensor subspace and their low-dimensional approximations are estimated.
\item The low-dimensional approximation tensors are used to fit the regression
model in Equation (\ref{eq:treg2}), and the parameters are estimated
via one of the methods described in Sections \ref{sec:cp} and \ref{sec:tucker}.
\item The image stream from the fielded system is projected onto the tensor
subspace estimated in step (iii), and its low-dimensional approximation
is also estimated. Next, the approximated tensor is input into the
regression model estimated in step (iv), and the TTF is predicted.
The RUL is obtained by subtracting the current observation time from
the predicted TTF. 
\end{enumerate}
Note that steps (i)-(iv) can be done offline. That is, given a training
dataset, we can construct \textit{truncated training datasets} with
images observed at time $\{t_{1},t_{2}\}$, $\{t_{1},t_{2},t_{3}\}$,
$\{t_{1},t_{2},t_{3},t_{4}\}$, $\ldots$, respectively. Regression
models are then estimated based on all the possible \textit{truncated
training datasets}. Once a new image is observed at say time $t_{n}$,
the appropriate regression model with images $\{t_{1},\ldots,t_{n}\}$
is chosen, and the RUL of the fielded system is estimated in step
(v). This approach enables real-time RUL predictions.

\section{Simulation study}

\label{sec:simulation}

In this section, we validate the effectiveness of our methodology
with the two types of decomposition approaches using simulated degradation
image streams. We assume the underlying physical degradation follows
a heat transfer process based on which simulated degradation image
streams are generated.

\subsection{Data generation}

\begin{figure}[!]
\centering \subfloat[Without noise]{\includegraphics[width=160mm]{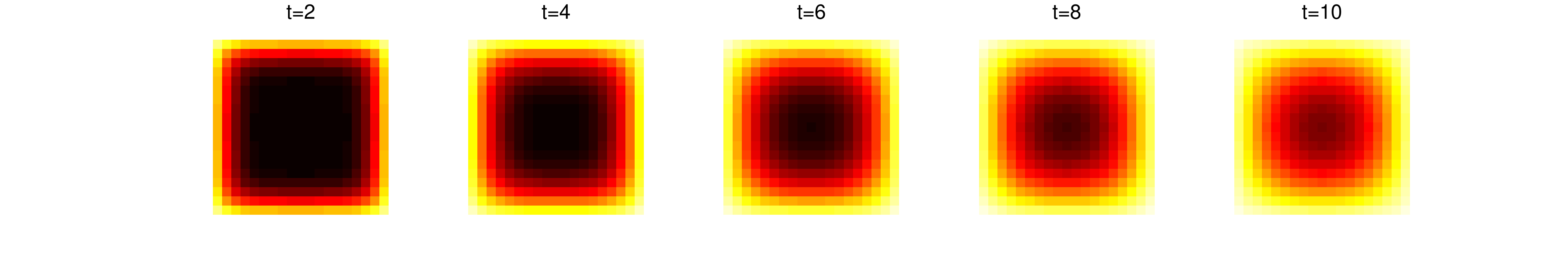}

}

\qquad{}\\
 \subfloat[With noise]{\includegraphics[width=160mm]{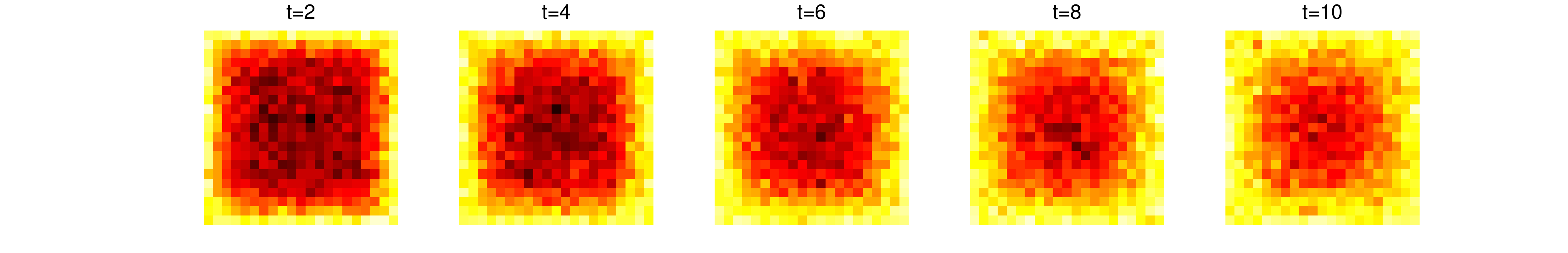}

}

\caption{Simulated degradation images based on heat transfer process.}
\label{fig:exdegimage} 
\end{figure}

Suppose for system $i$, the degradation image stream, denoted by
$\mathcal{S}_{i}(x,y,t)$, $i=1,\ldots,1000$, is generated from the
following heat transfer process: $\frac{\partial\mathcal{S}_{i}(x,y,t)}{\partial t}=\alpha_{i}(\frac{\partial^{2}\mathcal{S}_{i}}{\partial x^{2}}+\frac{\partial^{2}\mathcal{S}_{i}}{\partial y^{2}})$,
where $(x,y);0\leq x,y\leq0.05$ represents the location of each image
pixel, $\alpha_{i}$ is the thermal diffusivity coefficient for system
$i$ and is randomly generated from $uniform(0.5\times10^{-5},1.5\times10^{-5})$
and $t$ is the time frame. The initial and boundary conditions are
set such that $\mathcal{S}|_{t=1}=0$ and $\mathcal{S}|_{x=0}=\mathcal{S}|_{x=0.05}=\mathcal{S}|_{y=0}=\mathcal{S}|_{y=0.05}=1$.
At each time $t$, the image is recorded at locations $x=\frac{j}{n+1},y=\frac{k}{n+1},j,k=1,\ldots,n$,
resulting in an $n\times n$ matrix. Here we set $n=21$ and $t=1,\ldots,10$,
which leads to $10$ images of size $21\times21$ for each system
represented by a $21\times21\times10$ tensor. Finally, i.i.d noises
$\varepsilon\sim N(0,0.01)$ are added to each pixel. Example degradation
images observed at time $t=2,4,6,8,10$ from a simulated system are
shown in Figure \ref{fig:exdegimage}, in which (a) and (b) show images
without and with noise, respectively.

To simulate the TTF of each system two sets of coefficient tensors
are used. The first set, denoted by $\mathcal{B}_{C}$, is simulated
in the form of basis matrices with rank $2$ used in CP decomposition.
Specifically, three matrices, i.e., $\boldsymbol{B}_{C,1}\in\mathbb{R}^{21\times2},\boldsymbol{B}_{C,2}\in\mathbb{R}^{21\times2},\boldsymbol{B}_{C,3}\in\mathbb{R}^{10\times2}$
are generated. To induce sparsity, we randomly set half of elements
of each matrix to be $0$. The values of the remaining $50\%$ elements
are randomly generated from a uniform distribution $unif(-1,1)$.
The TTF, denoted by $y_{C,i}$, is generated by using $y_{C,i}=\langle vec(\mathcal{B}_{C}),vec(\mathcal{S}_{i})\rangle+\sigma\epsilon_{i}$,
where $vec({\mathcal{B}_{C}})=(\boldsymbol{B}_{C,3}\odot\boldsymbol{B}_{C,2}\odot\boldsymbol{B}_{C,1})\boldsymbol{1}_{2}$,
$\epsilon_{i}$ follows a standard smallest extreme value distribution
$\text{SEV}(0,1)$ and $\sigma$ is $5\%$ times the standard deviation
of the location parameter, i.e., $\langle vec(\mathcal{B}_{C}),vec(\mathcal{S}_{i})\rangle$.

The second set, denoted by $\mathcal{B}_{T}$, is simulated in the
form of core and factor matrices with rank $(2,1,2)$ used in Tucker
decomposition. Specifically, a core tensor $\mathcal{G}_{T}\in\mathbb{R}^{2\times1\times2}$
and three factor matrices $\boldsymbol{B}_{T,1}\in\mathbb{R}^{21\times2},\boldsymbol{B}_{T,2}\in\mathbb{R}^{21\times1},\boldsymbol{B}_{T,3}\in\mathbb{R}^{10\times2}$
are generated. All the elements of the core tensor $\mathcal{G}_{T}$
are set to $1$. Furthermore, half of elements of matrices $\boldsymbol{B}_{T,1},\boldsymbol{B}_{T,2},\boldsymbol{B}_{T,3}$
are randomly set to $0$ and the remaining elements are randomly generated
from $unif(-1,1)$. The TTF, $y_{T,i}$, is generated via $y_{T,i}=\langle vec(\mathcal{B}_{T}),vec(\mathcal{S}_{i})\rangle+\sigma\epsilon_{i}$,
where $vec({\mathcal{B}_{T}})=\mathcal{G}_{T}\times_{1}\boldsymbol{B}_{T,1}\times_{2}\boldsymbol{B}_{T,2}\times_{3}\boldsymbol{B}_{T,3}$,
$\epsilon_{i}$ follows a standard smallest extreme value distribution
$\text{SEV}(0,1)$ and $\sigma$ is $5\%$ times the standard deviation
of the location parameter, i.e., $\langle vec(\mathcal{B}_{T}),vec(\mathcal{S}_{i})\rangle$.

In the following two subsections, we use the simulated degradation
data to validate the proposed methodology. We first study the performance
of the BIC criterion and our heuristic rank selection method in identifying
the correct LLS distribution (i.e., SEV) as well as the right rank.
Then, the prediction capability of our prognostic model is evaluated
at different life percentiles of simulated systems. We randomly select
$500$ of the simulated systems for training and the remaining $500$
systems for test.

\subsection{Model and rank selection}

We first apply  CP-based tensor regression  in Equation (\ref{eq:treg3})
to the training dataset, $\{y_{C,i},\mathcal{S}_{i}\}_{i=1}^{500}$,
and use Algorithm 1 to estimate the model parameters for different
ranks and for four LLS distributions, namely, \textit{normal, SEV},
\textit{lognormal} and \textit{Weibull}. The BIC value is then computed
for each distribution and rank combination as discussed in Section
\ref{sec:cp}. As pointed out earlier, the block relaxation method
in Algorithm 1 only guarantees a local optimum and hence, we shall
run the algorithm 10 times using randomized initializations and record
the smallest BIC. The BIC values for all combinations are reported
in Table \ref{lb:CPBIC}. From Table \ref{lb:CPBIC}, it can be seen
that the smallest BIC value is -1535.3, which belongs to \textit{SEV}
distribution with rank $R=2$. This coincides with the rank and the
distribution we used to generate the data.

\begin{table}[h]
\centering{}\caption{BIC values for CP-based tensor regression.}
\label{lb:CPBIC} %
\begin{tabular}{cccccccc}
\hline 
Rank  & R=1  & R=2  & R=3  & R=4  & R=5  & R=6  & R=7\tabularnewline
\hline 
SEV  & 620.5  & \textbf{-1535.3}  & -1383.4  & -1232.7  & -1122.9  & -1014.4  & -805.9\tabularnewline
Normal  & 550.0  & -1422.6  & -1273.6  & -1153.2  & -1064.7  & -1013.2  & -1114.0\tabularnewline
Weibull  & 618.1  & -643.6  & -472.6  & -301.9  & -180.5  & -103.5  & -54.0\tabularnewline
Lognormal  & 610.5  & -336.3  & -187.7  & -75.5  & 9.6  & 67.4  & 74.3\tabularnewline
\hline 
\end{tabular}
\end{table}

Similarly, the Tucker-based tensor regression model in Equation (\ref{eq:tuckerreg1})
is applied to the training dataset, $\{y_{T,i},\mathcal{S}_{i}\}_{i=1}^{500}$
and Algorithm 2 (see Section \ref{sec:tucker}) is used to estimate
the parameters. A total of $27$ different rank combinations are tested
under four distributions, \textit{normal, SEV}, \emph{lognormal} and
\emph{Weibull}. Again, for each distribution-rank combination, Algorithm
2 is run with $10$ randomized initializations, and the smallest BIC
value is reported in Table \ref{lb:TuckerBIC} .

\begin{table}[h]\small
\centering{}\caption{BIC values for Tucker-based tensor regression.}
\label{lb:TuckerBIC} %
\begin{tabular}{llllllllll}
\hline
\noalign{\vskip\doublerulesep}
Rank  & (1,1,1)  & (1,1,2) & (1,1,3)  & (1,2,1) & (1,2,2) & (1,2,3) & (1,3,1) & (1,3,2) & (1,3,3)\tabularnewline[\doublerulesep]
\hline
\noalign{\vskip\doublerulesep}
SEV  & -163.3  & -113.2  & -75.8 & -44.8  & -59.0 & -15.7 & 61.0 & 52.6 & 29.6 \tabularnewline[\doublerulesep]
\noalign{\vskip\doublerulesep}
Normal  & -199.0 & -149.3 & -112.0 & -81.0 & -82.8 & -39.3 & 24.8 & 28.9 & 15.5 \tabularnewline[\doublerulesep]
\noalign{\vskip\doublerulesep}
Weibull & -73.9 & -24.4 & 13.0 & 44.1 & 35.9 & 79.2 & 149.6 & 147.4 & 133.5\tabularnewline[\doublerulesep]
\noalign{\vskip\doublerulesep}
Lognormal & -83.7 & -33.9 & 3.4 & 34.4 & 28.6 & 71.6 & 140.0 & 140.2 & 141.9\tabularnewline[\doublerulesep]
\hline
\noalign{\vskip\doublerulesep}
Rank  & (2,1,1)  & (2,1,2) & (2,1,3)  & (2,2,1) & (2,2,2) & (2,2,3) & (2,3,1) & (2,3,2) & (2,3,3)\tabularnewline[\doublerulesep]
\hline
\noalign{\vskip\doublerulesep}
SEV  & -44.8 & \textbf{-1313.5} & -1269.8  & -16.1  & -1212.7 & -1202.9 & 95.4 & -1115.0 & -1106.5 \tabularnewline[\doublerulesep]
\noalign{\vskip\doublerulesep}
Normal  & -80.9 & -1259.1 & -1215.6 & -22.9 & -1149.7 & -1130.8 & 89.3 & -1048.6 & -1028.2 \tabularnewline[\doublerulesep]
\noalign{\vskip\doublerulesep}
Weibull & 44.1 & -733.8 & -690.2 & 66.1 & -633.2 & -607.1 & 178.1 & -543.5 & -508.1\tabularnewline[\doublerulesep]
\noalign{\vskip\doublerulesep}
Lognormal & 34.4 & -497.8 & -454.3 & 85.2- & -402.7 & -394.4 & 197.2 & -306.2 & -292.6\tabularnewline[\doublerulesep]
\hline
\noalign{\vskip\doublerulesep}
Rank  & (3,1,1)  & (3,1,2) & (3,1,3)  & (3,2,1) & (3,2,2) & (3,2,3) & (3,3,1) & (3,3,2) & (3,3,3)\tabularnewline[\doublerulesep]
\hline
\noalign{\vskip\doublerulesep}
SEV  & 60.7  & -1201.8 & -1224.9  & 95.5  & -1156.4 & -1164.0 & 113.0 & -1071.4 & -1074.4 \tabularnewline[\doublerulesep]
\noalign{\vskip\doublerulesep}
Normal  & 24.9 & -1147.2 & -1153.2 & 88.8 & -1093.2 & -1082.6 & 129.4 & -1009.0 & -999.2 \tabularnewline[\doublerulesep]
\noalign{\vskip\doublerulesep}
Weibull & 149.7 & -621.9 & -613.1 & 177.8 & -572.3 & -539.2 & 205.6 & -488.5 & -468.2\tabularnewline[\doublerulesep]
\noalign{\vskip\doublerulesep}
Lognormal & 139.9 & -385.9 & -391.0 & 197.5 & -337.9 & -331.4 & 238.5 & -252.0 & -262.3\tabularnewline[\doublerulesep]
\hline
\end{tabular}
\end{table}

Table \ref{lb:TuckerBIC} indicates that the smallest BIC value (-1313.5)
is associated with the \textit{SEV} distribution with rank $R=(2,1,2)$,
which again matches the rank and the distribution that was used to
generate the data. Therefore, we can conclude that the BIC criterion
is effective in selecting an appropriate distribution as well as the
correct rank of the tensors in the LLS regression. In Table \ref{lb:AutoRank},
we also report the results of the heuristic rank selection method
for Tucker. It can be seen from Table \ref{lb:AutoRank} that the
heuristic rank selection method selects rank $R=(1,1,1)$ under \emph{normal}
and \emph{lognormal} distributions, while selects rank $R=(2,2,2)$
under\emph{ SEV} distribution and $R=(1,2,2)$ under\emph{ Weibull}
distribution. The smallest BIC values (-1212.3) is achieved under
\emph{SEV} distribution with rank $R=(2,2,2)$, which is close to
the actual rank.

\begin{table}[h]
\centering{}\caption{Distribution and rank selection results by using heuristic rank selection
method.}
\label{lb:AutoRank} %
\begin{tabular}{ccc}
\hline 
LLS Distribution  & Rank  & BIC\tabularnewline
\hline 
SEV  & (2,2,2)  & \textbf{-1212.3}\tabularnewline
Normal  & (1,1,1)  & -199.0\tabularnewline
Weibull  & (1,2,2)  & 36.1\tabularnewline
Lognormal  & (1,1,1)  & -83.7\tabularnewline
\hline 
\end{tabular}
\end{table}

\subsection{RUL prediction results}

\label{subsec:pred}

\begin{figure}[htb!]
\centering{}\centering \includegraphics[width=0.6\textwidth]{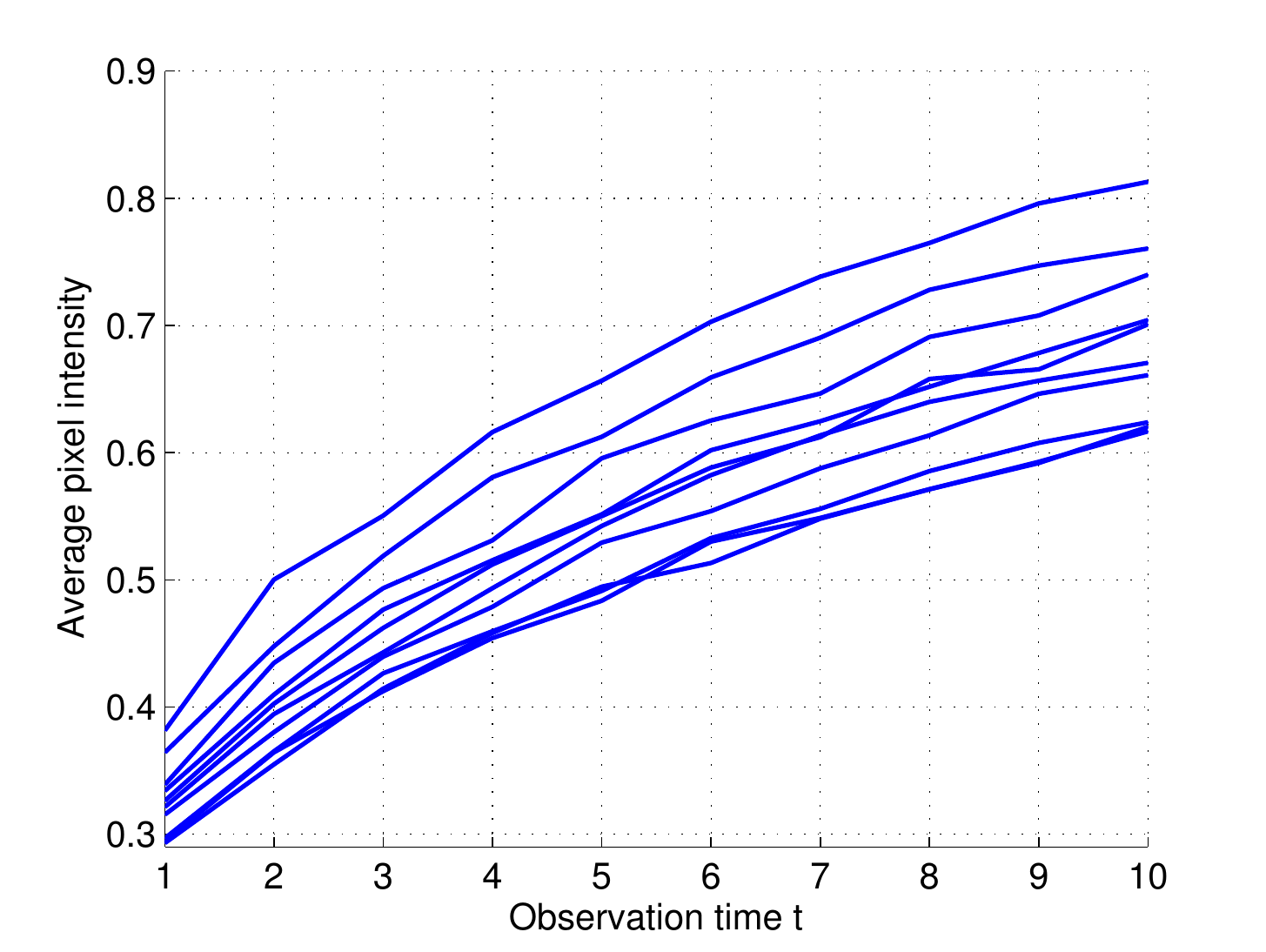}
\caption{A sample of time-series-based degradation signals.}
\label{fig:simavetemp} 
\end{figure}

We evaluate the prediction accuracy of our prognostic model and compare
its performance to a benchmark that uses functional principal components
analysis (FPCA) to model the overall image intensity, which we designate
as ``FPCA\char`\"{}. For the benchmark, we first transform the degradation
image stream of each system into a time-series signal by taking the
average intensity of each observed image. A sample of transformed
time-series degradation signals is shown in Figure \ref{fig:simavetemp}.
Next, FPCA is applied to the time-series signals to extract features.
FPCA is a popular functional data analysis technique that identifies
the important sources of patterns and variations among functional
data (time-series signals in our case)\citep{Ramsay2005}. The time-series
signals are projected to a low-dimensional feature space spanned by
the eigen-functions of the signals' covariance function and provides
fused features called FPC-scores. Finally, FPC-scores are regressed
against the TTF by using LLS regression  under \textit{SEV} distribution.
More details about this FPCA prognostic model can be found in \citep{Fang2015}.

The CP-based tensor regression model with rank $R=2$ under \textit{SEV}
distribution is applied to the training dataset to estimate the model.
Next, the trained model is used to predict the TTFs of the test dataset
using their degradation-based image observations. Prediction errors
are calculated using the following expression:

\begin{equation}
\text{Prediction Error}=\frac{|\text{Estimated Lifetime-Real Lifetime}|}{\text{Real Lifetime}}\label{eq:er}
\end{equation}

\begin{figure}[htb!]
\centering{}\centering \includegraphics[width=0.6\textwidth]{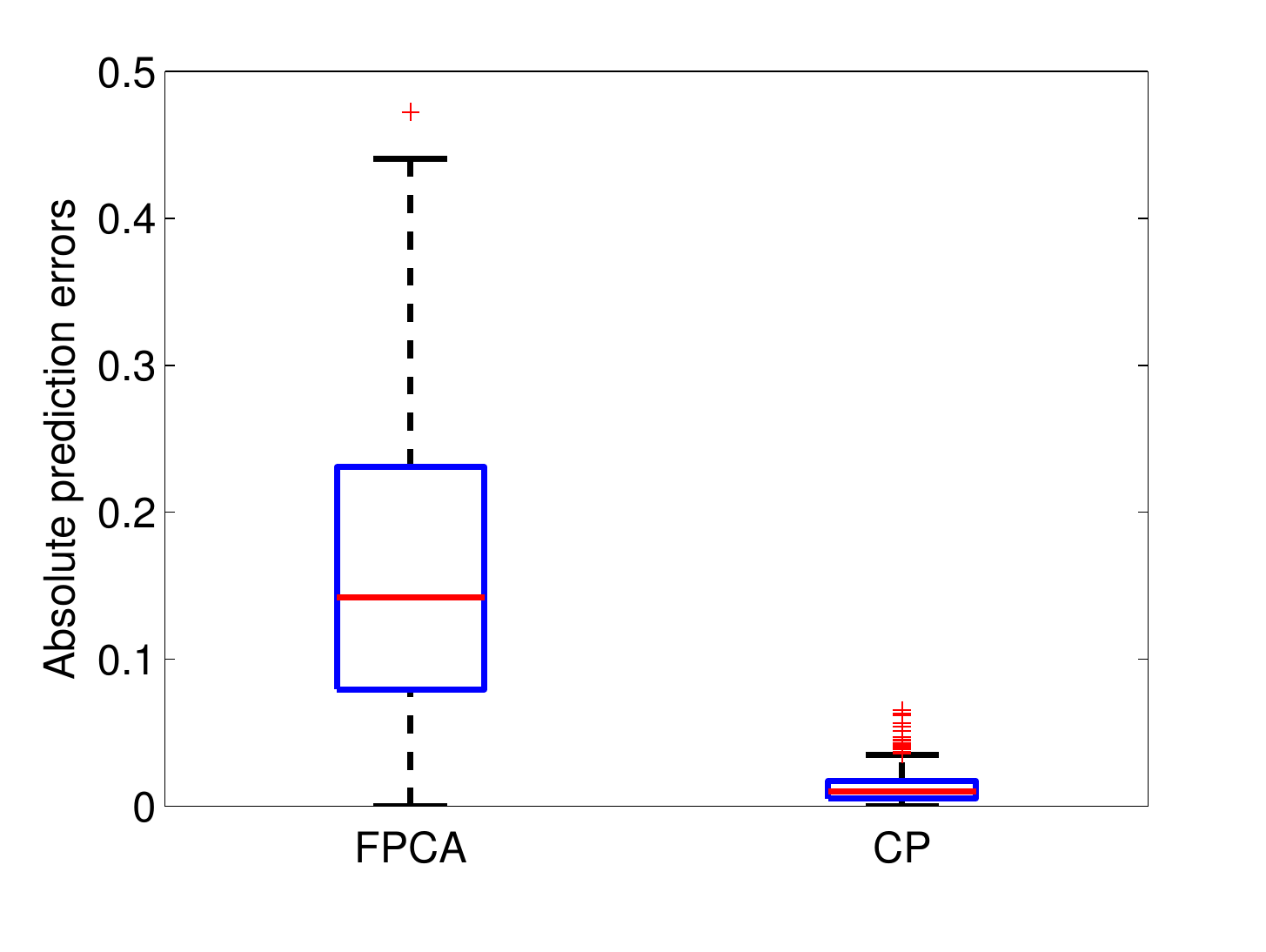}
\caption{Prediction errors of CP-based regression under SEV distribution.}
\label{fig:CPPrediction} 
\end{figure}

\begin{figure}[htb!]
\centering{}\centering \includegraphics[width=0.6\textwidth]{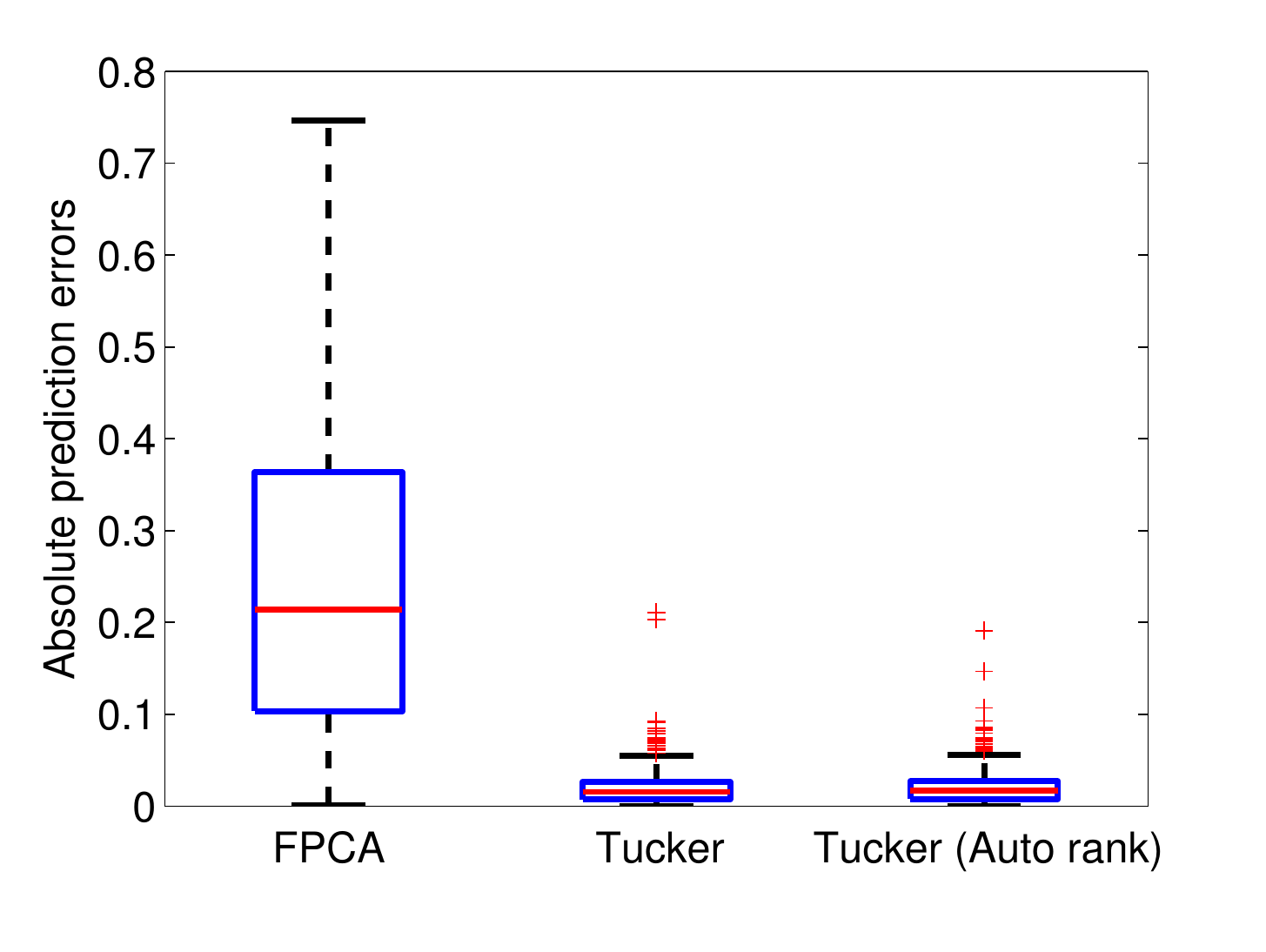}
\caption{Prediction errors of Tucker-based regression under SEV distribution.}
\label{fig:TuckerPrediction} 
\end{figure}

Two Tucker-based tensor regression models, one with rank $R=(2,1,2)$
selected by BIC and another with rank $R=(2,2,2)$ selected by heuristic
rank selection method, are applied to the data under \textit{SEV}
distribution. The prediction errors of CP-based and Tucker-based tensor
regression are reported in Figure \ref{fig:CPPrediction} and \ref{fig:TuckerPrediction},
respectively.

Figure \ref{fig:CPPrediction} shows that the accuracy of the CP-based
model is significantly better than the FPCA benchmark, where the mean
prediction error  is $15\%$ and $1\%$ for FPCA and our CP-based
tensor regression, respectively. The variance of the prediction errors
for FPCA is also much larger than that of CP-based tensor regression.
Similar findings can be seen in Figure \ref{fig:TuckerPrediction},
where the mean prediction error of FPCA is 20\%, and it is around
1\% for Tucker-based tensor regression. The variance of the prediction
errors for Tucker-based tensor regression is again much smaller than
that of FPCA. The performance of the FPCA benchmark relative to our
methodology highlights the importance of accounting for the spatial-temporal
structure of  image streams. The transformation of image streams to
time-series signals ignores the spatial structure of images resulting
in significant loss of information, and thus, compromising the prediction
accuracy.

Figure \ref{fig:TuckerPrediction} highlights the insignificant difference
in prediction accuracy of the two proposed rank selection methods,
BIC and heuristic rank selection (denoted ``Tucker (Auto rank)''
in the figure). This result further validates the effectiveness of
our automatic rank selection method.

\section{Case study: Degradation image streams from rotating machinery}

\label{sec:case}

In this section, we validate the effectiveness of our methodology
using degradation image streams obtained from a rotating machinery.
The experimental test bed, which was described in detail in \citet{Gebraeel2009},
is designed to perform accelerated degradation tests on rolling element
thrust bearings. The test bearings are run from a brand new state
until failure. Vibration sensors are used to monitor the health of
the rotating machinery. Failure is defined once the amplitude of defective
vibration frequencies crosses a pre-specified threshold based on ISO
standards for machine vibration. Meanwhile, infrared images that capture
temperature signatures of the degraded component throughout the degradation
test are acquired using an FLIR T300 infrared camera. Infrared images
with $40\times20$ pixels are stored every $10$ seconds. Four different
experiments were run to failure. The resulting degradation-based image
streams contained $375$, $611$, $827$ and $1,478$ images, respectively.

\begin{figure}[h]
\centering \vspace{-1em}
 \subfloat[t=1]{\includegraphics[width=20mm]{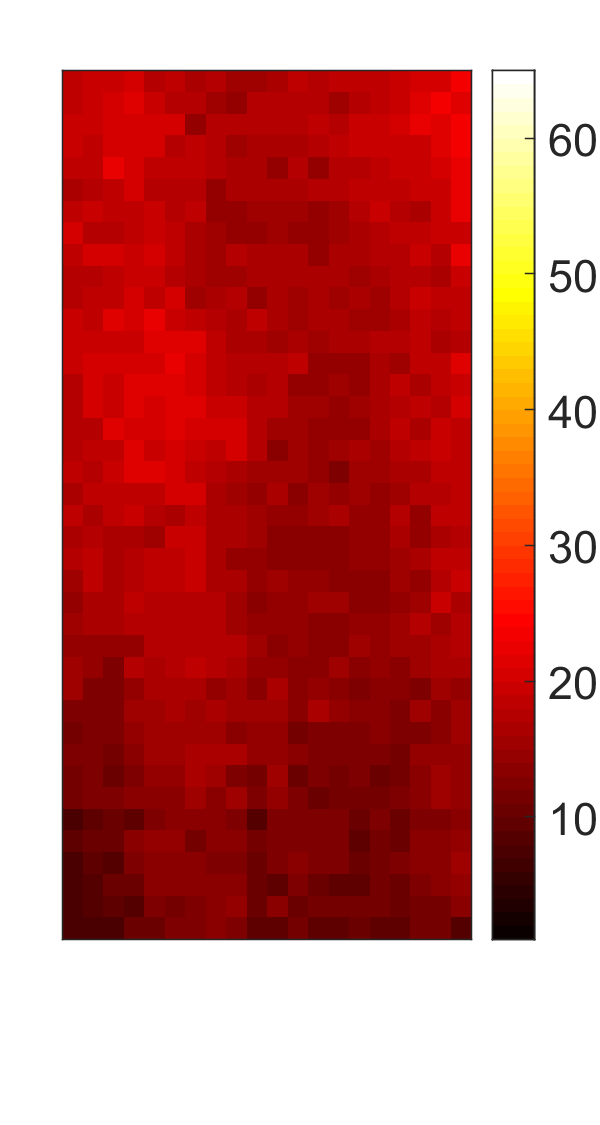}

} \qquad{}\subfloat[t=2]{\includegraphics[width=20mm]{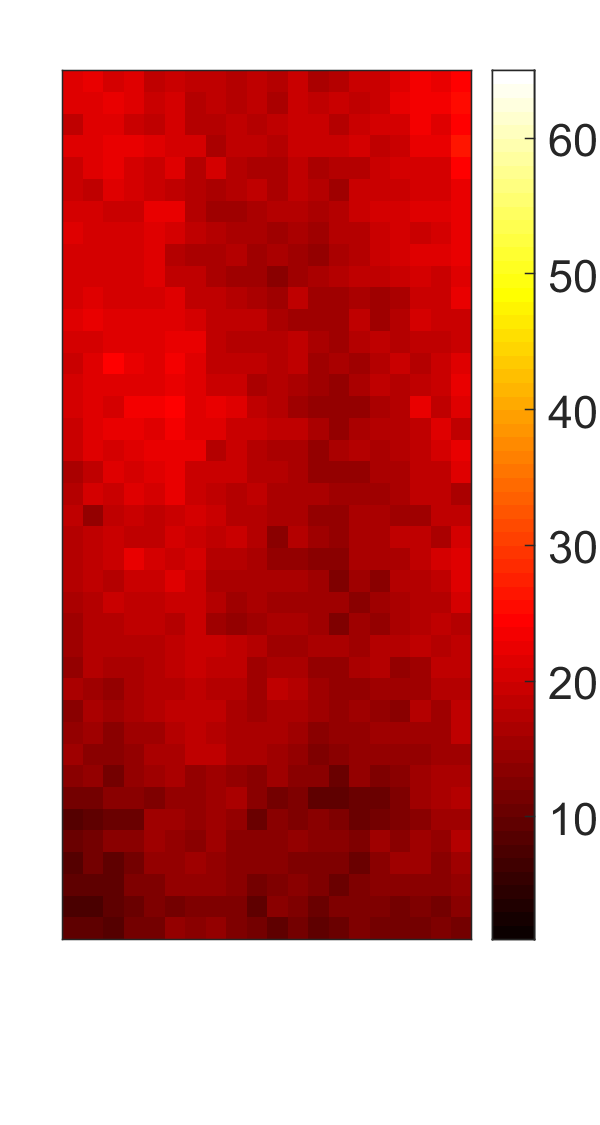}

} \qquad{}\subfloat[t=3]{\includegraphics[width=20mm]{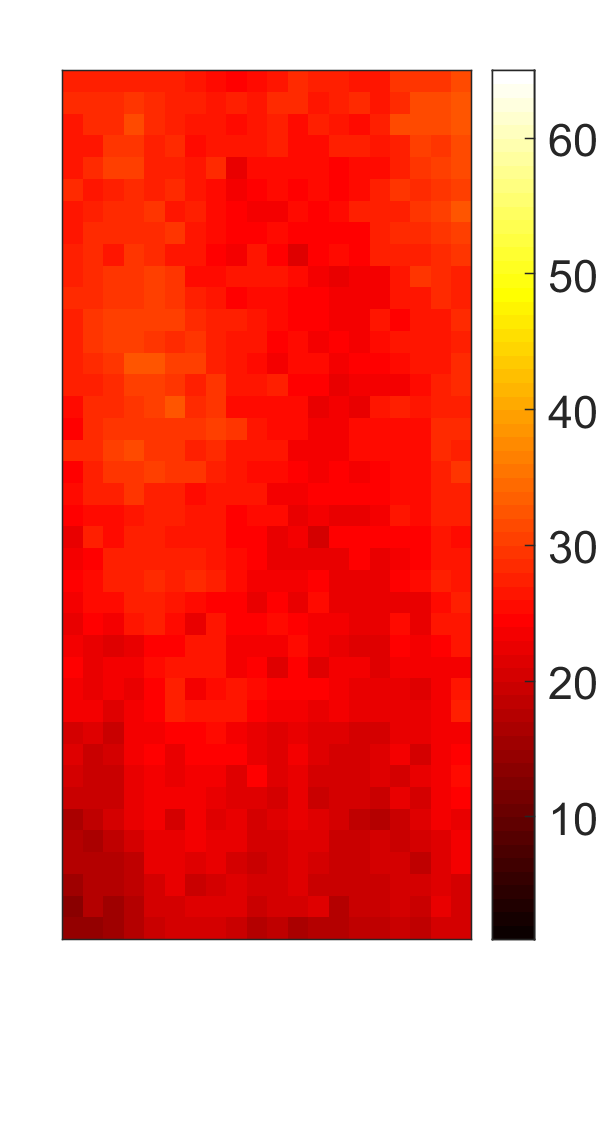}

} \qquad{}\subfloat[t=4]{\includegraphics[width=20mm]{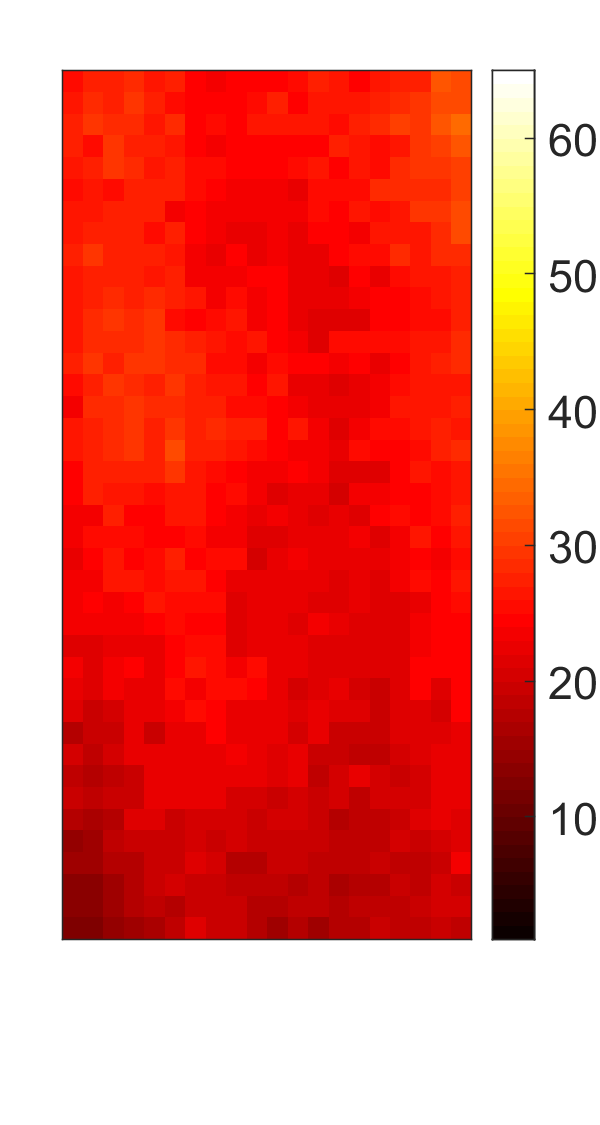}

} \qquad{}\subfloat[t=5]{\includegraphics[width=20mm]{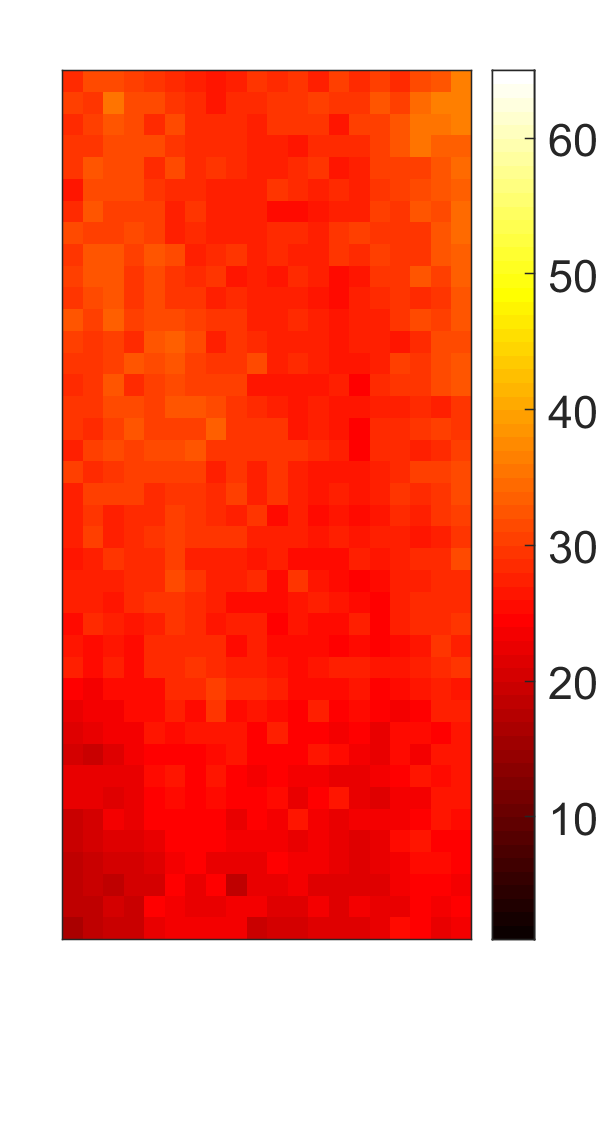}

}\vspace{-1em}

\subfloat[t=6]{\includegraphics[width=20mm]{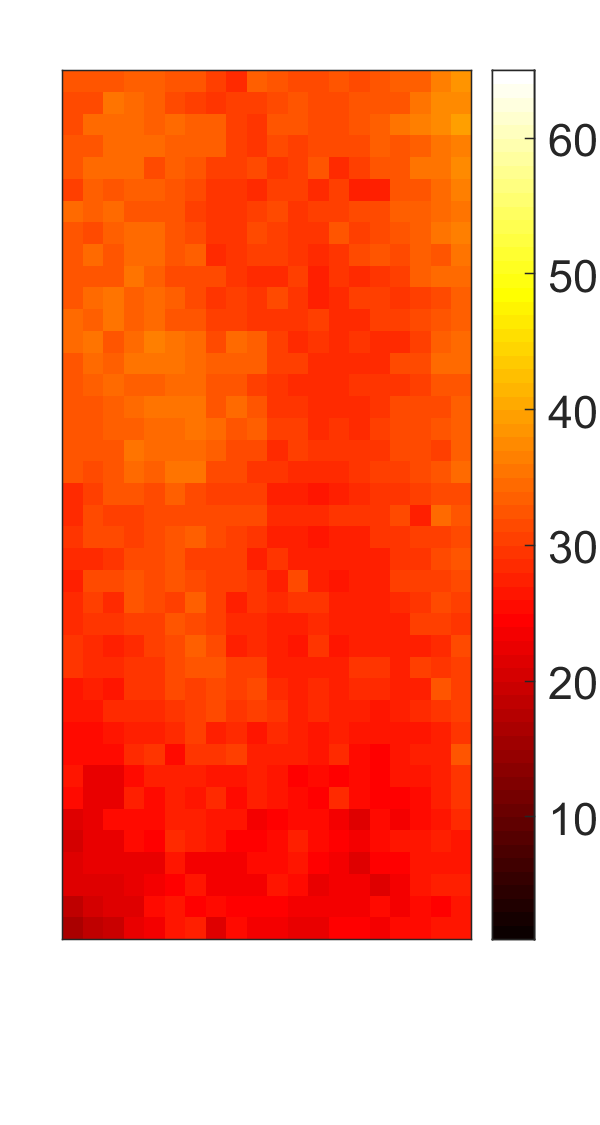}

} \qquad{}\subfloat[t=7]{\includegraphics[width=20mm]{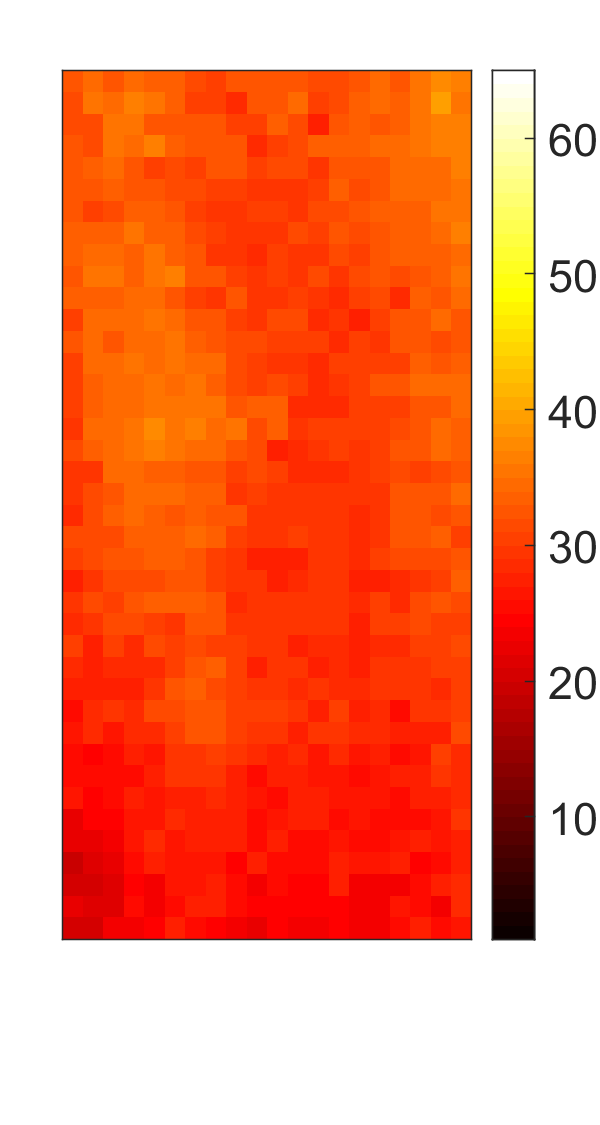}

} \qquad{}\subfloat[t=8]{\includegraphics[width=20mm]{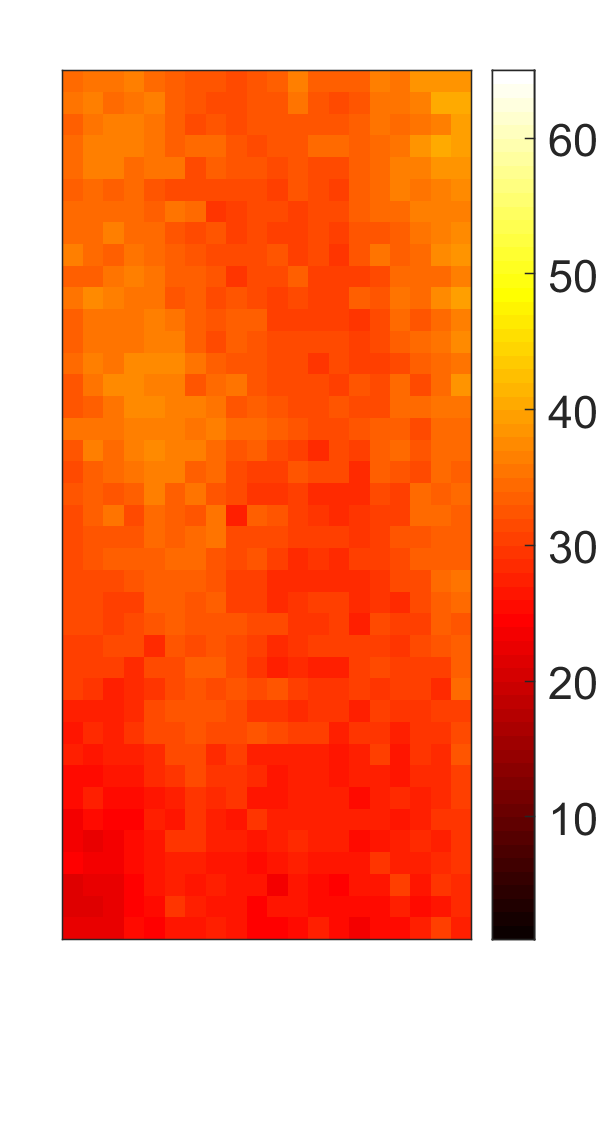}

} \qquad{}\subfloat[t=9]{\includegraphics[width=20mm]{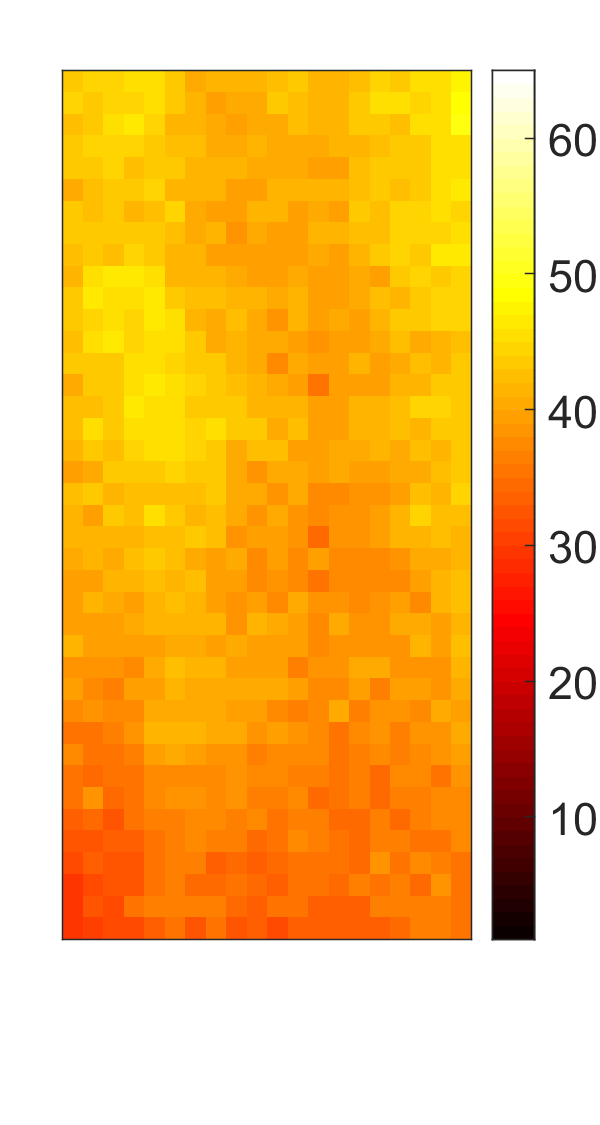}

} \qquad{}\subfloat[t=10]{\includegraphics[width=20mm]{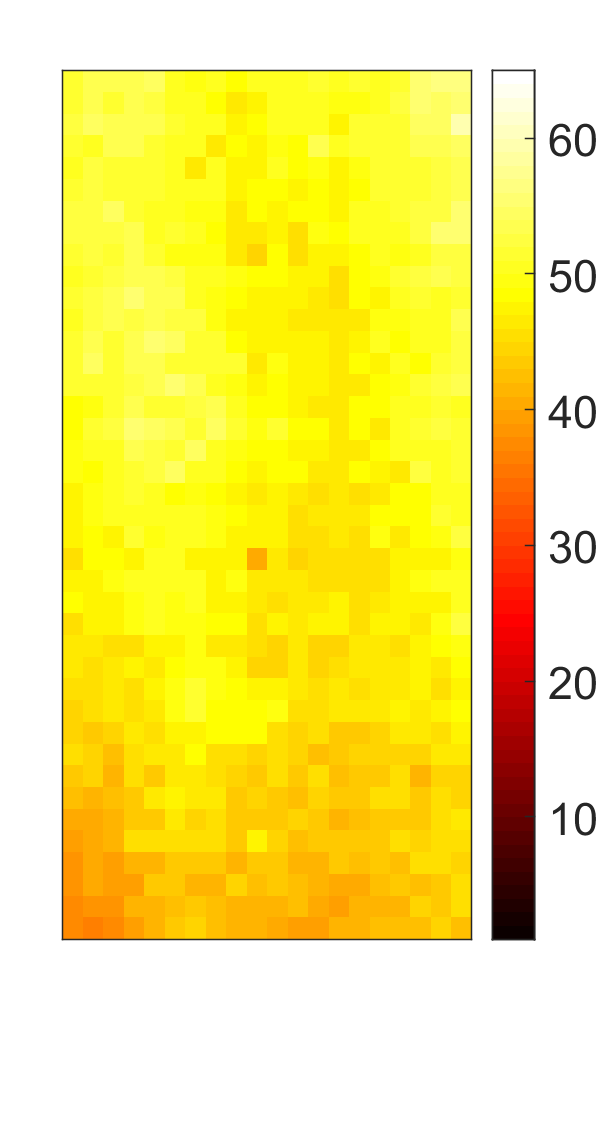}

}

\caption{An illustration of one infrared degradation image stream.}
\label{fig:exp} 
\end{figure}

Due to the high cost of running degradation experiments, additional
degradation image streams were generated by resampling from the original
image database obtained from the four experiments discussed earlier.
In total 284 image data streams were generated. As an illustration,
a sequence of images obtained at different (ordered) time periods
are shown in Figure \ref{fig:exp}.

\subsection{Model selection}

In this section, we discuss how to select an appropriate LLS tensor
regression model for our dataset. This is achieved by applying different
LLS tensor regression candidate models to a training set consisting
of multiple image data streams. The model with the smallest BIC is
selected as the best candidate model.

To account for the variability in the length of the image streams
(as illustrated earlier in Section \ref{sec:updating}), we generate
multiple subsamples based on different TTFs. Specifically, we sort
the TTFs in ascending order such that $TTF_{1}<TTF_{2}<\cdots<TTF_{n}$,
where $n\leq284$ is the number of unique TTFs (or equivalent the
number of subsamples). Next, we define subsample $i$ as the systems
whose TTFs are greater than or equal to $TTF_{i}$, for $i=1,\ldots,n$.
For example, subsample $1$ includes all the $284$ image streams,
and subsample $2$ includes all the image streams excluding the ones
with the smallest TTF, and so forth. Third, each subsample is truncated
by only keeping images observed on time domain $[0,TTF_{i}]$ epochs.
By doing so, we ensure that all the image streams in a subsample have
the same dimensionality. This is important when applying the LLS tensor
regression model. After truncation, the following steps are applied
to select the best candidate regression model: 
\begin{itemize}
\item \emph{Step 1: Dimension reduction}. MPCA is applied to each subsample
$i$ (truncated image stream). The fraction-of-variance-explained,
which is used to select the number of multilinear principal components
(see \citet{Lu2008} for details), is set to be $0.95$. Using this
criterion, a low-dimensional tensor is extracted from each image stream
(or each system). 
\item \emph{Step 2: Fitting LLS model}. The low-dimensional tensors extracted
from Step 1 are regressed against TTFs using an LLS regression model.
Similar to the Simulation study, we evaluate four types of distributions:
\emph{normal}, \emph{lognormal}, \emph{SEV} and \emph{Weibull}. Tucker-based
estimation method with heuristic rank selection is used for parameter
estimation. 
\item \emph{Step 3: Comparing BIC values.} BIC values are then computed
for each of the four fitted models. The model with the smallest BIC
is selected as the most appropriate one for the subsample. 
\item \emph{Step 4: Distribution selection.} Steps 1, 2, and 3 are applied
to all the subsamples. The distribution with the highest selected
frequency is considered as the best candidate distribution. 
\end{itemize}
After applying the aforementioned selection procedures to all the
subsamples, we summarize the percentage of times each distribution
was selected. Table \ref{lb:casetucker} summarizes these results
and shows that the \emph{Weibull} distribution was selected on average
$74.4\%$ while the lognormal was selected $25.6\%$ of the time.
We expect to have some overlap in the models that have been selected
because for specific parameter values, different distributions may
exhibit reasonable fits for the same data sets. In our case, it is
clear that the \emph{Weibull} distribution dominates most of the selections
and will therefore be considered as the suitable distribution for
this data set.

\begin{table}[h]
\centering{}\caption{Distribution selection results.}
\label{lb:casetucker} %
\begin{tabular}{ccccc}
\hline 
LLS Distribution  & Normal  & Lognormal  & SEV  & Weibull \tabularnewline
\hline 
Selection (\%)  & $0\%$  & $25.6\%$  & $0\%$  & $74.4\%$ \tabularnewline
\hline 
\end{tabular}
\end{table}

\subsection{Performance Evaluation}

\begin{figure}[htb!]
\centering \includegraphics[width=0.6\textwidth]{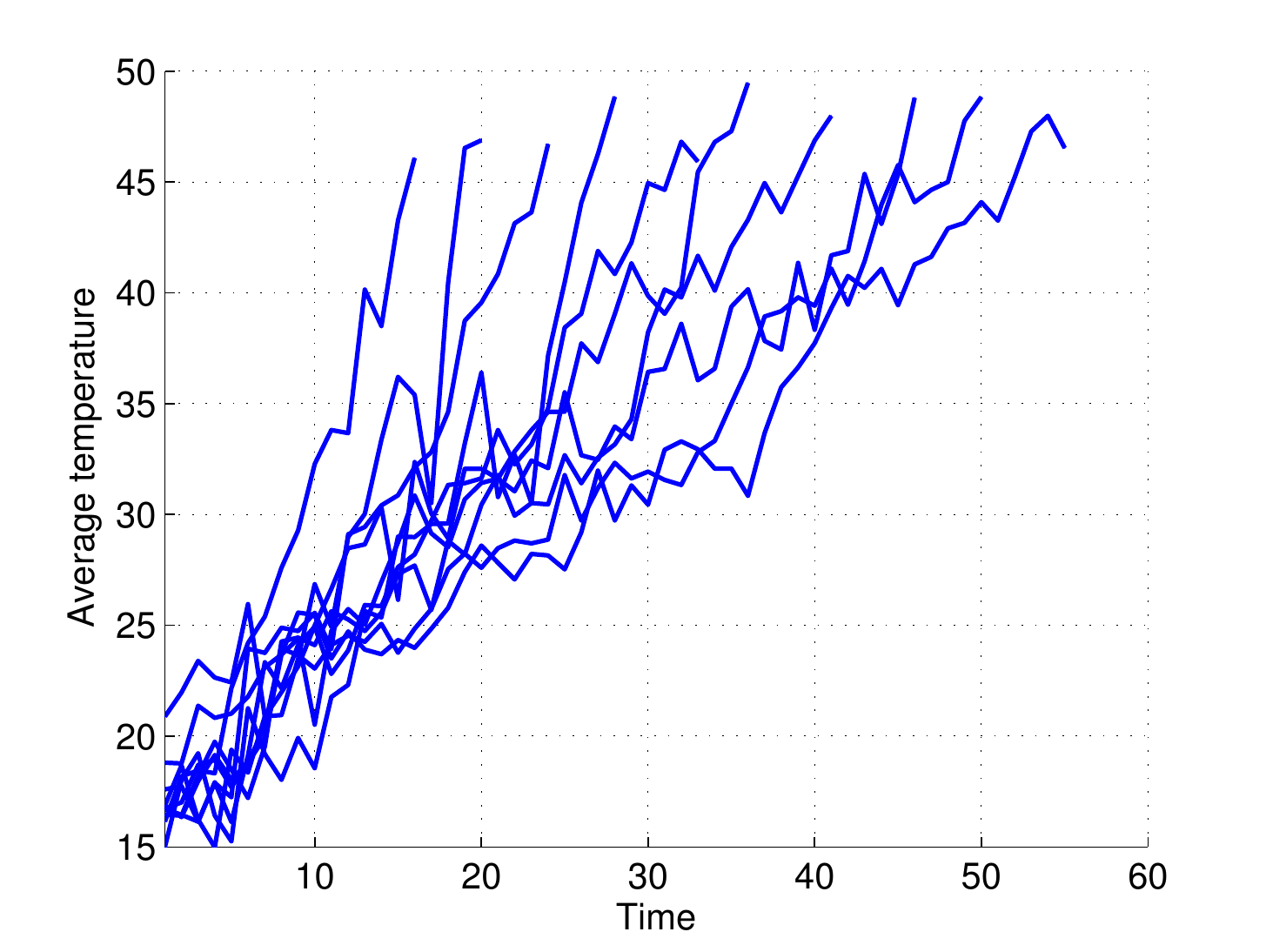} \caption{A sample of transformed time-series signals.}
\label{fig:avetemp} 
\end{figure}

The Weibull tensor regression model is chosen for evaluating the accuracy
of predicting lifetime. Similar to the simulation study in Section
\ref{sec:simulation}, we compare the performance of our methods with
the FPCA approach, denoted by ``FPCA''. Time-series degradation
signals corresponding to the infrared images of the experimental test
bed were obtained in a similar manner to what was discussed in Section
\ref{sec:simulation}. Figure \ref{fig:avetemp} shows a sample of
these transformed time-series signals.

\begin{figure}[h]
\subfloat[mean]{\begin{centering}
\includegraphics[width=0.49\textwidth]{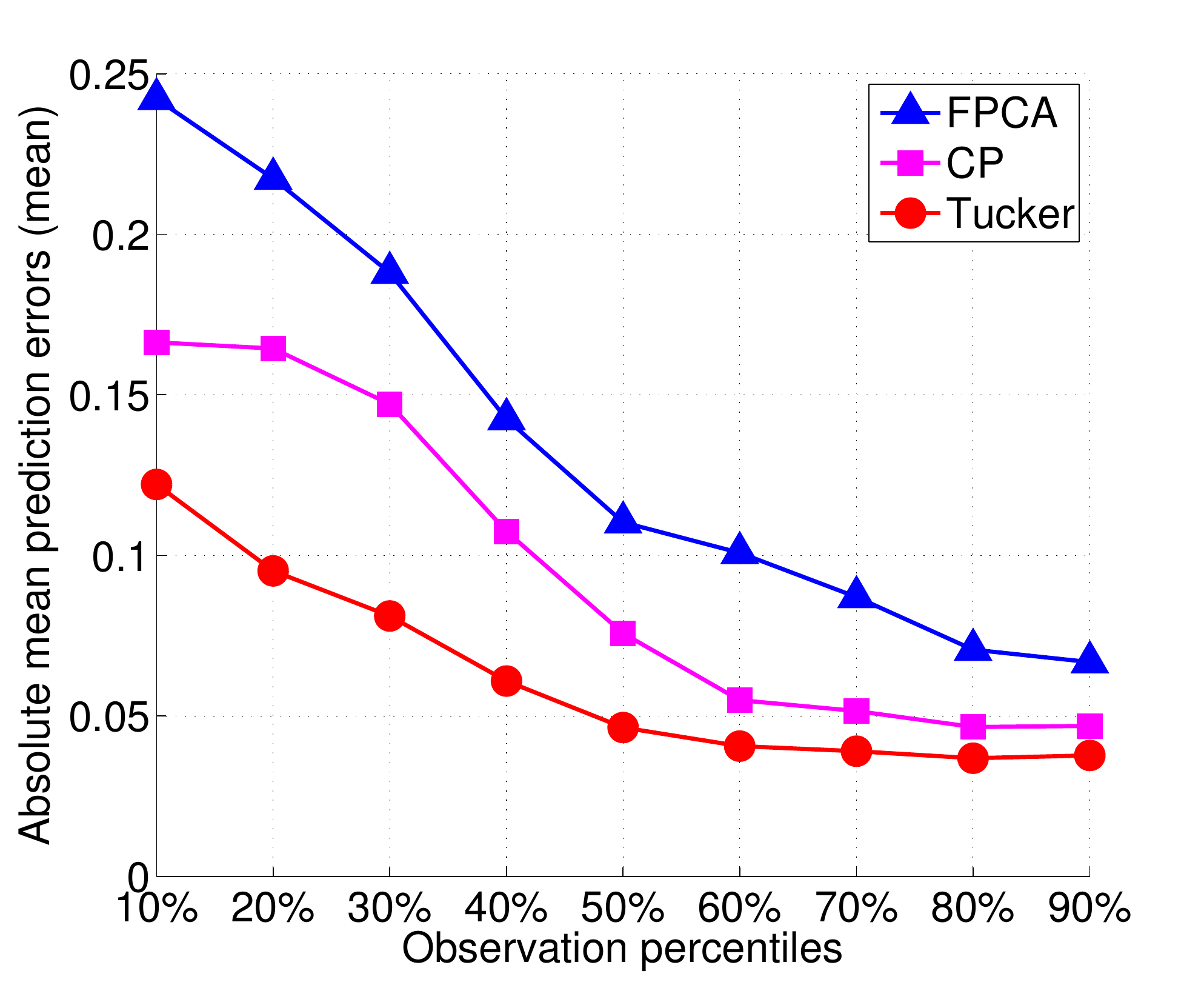}
\par\end{centering}
}\hfill{}\subfloat[variance]{\begin{centering}
\includegraphics[width=0.49\textwidth]{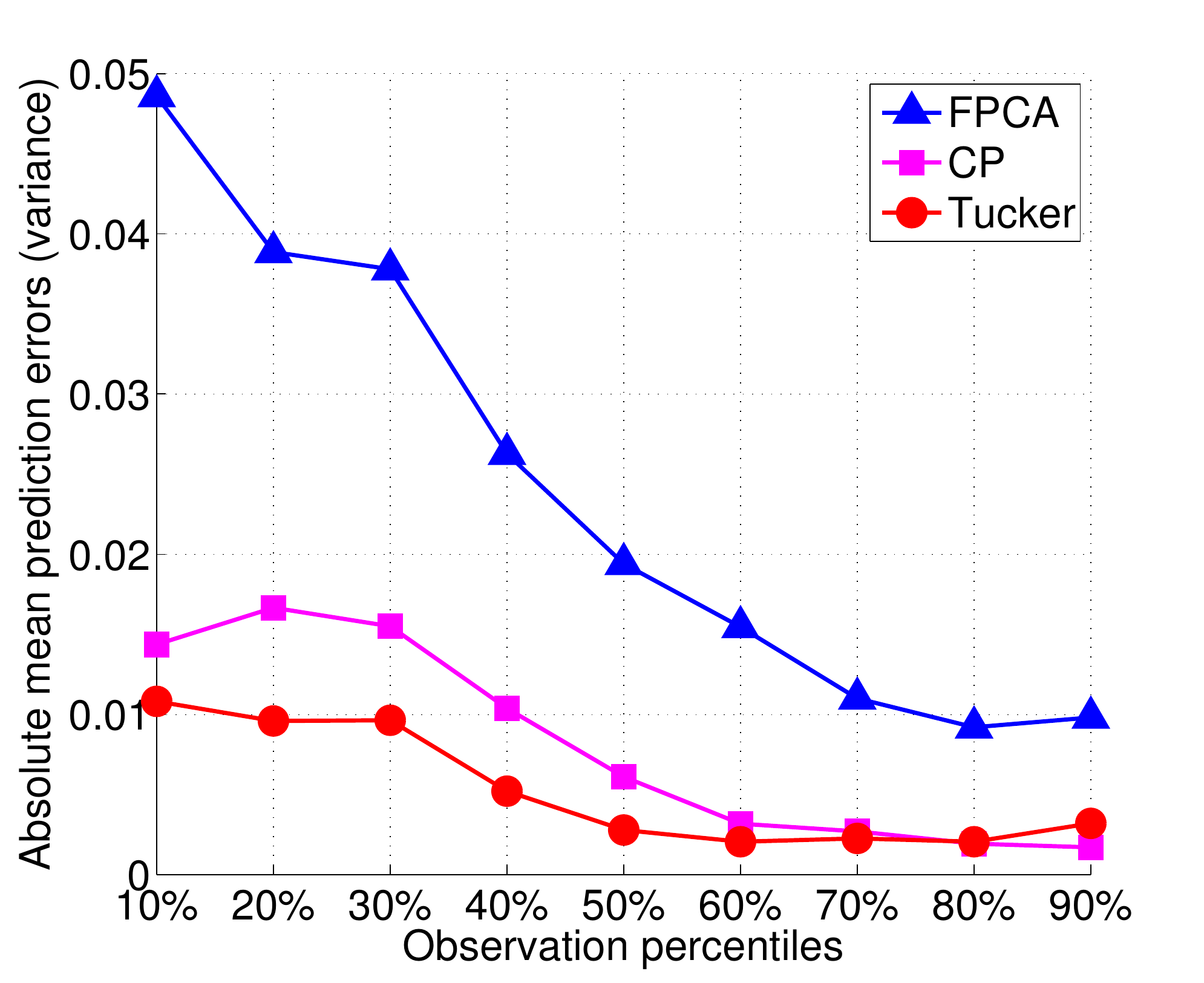}
\par\end{centering}
}

\caption{The mean and variance of the absolute prediction errors.}
\label{fig:er} 
\end{figure}

The accuracy and precision of the predictions made by the proposed
model as well as the FPCA model are evaluated using a leave-one-out
cross-validation study. For each validation, $283$ systems are used
for training and the remaining one system is used for testing. The
RULs of the test system are predicted at each time epoch. The time-varying
regression framework presented in Section \ref{sec:updating} is used
to enable the integration of newly observed image data (from the test
data). The prediction errors are computed using Equation (\ref{eq:er}).
We report the mean and variance of the absolute prediction errors
in Figure \ref{fig:er} where $10\%$ represents prediction errors
evaluated at life percentiles in the interval of $(5\%,15\%]$, $20\%$
for the interval of $(15\%,25\%]$, etc.

Figure \ref{fig:er} indicates that all the three methodologies have
smaller prediction errors at higher observation percentiles. This
is because at higher observation percentiles more degradation-based
image data has been observed, which provide more information about
the underlying physical degradation process. This results in better
prediction accuracy. Figure \ref{fig:er} also shows that the proposed
CP-based and Tucker-based regression models outperform the FPCA model
in terms of mean and variance of the absolute prediction errors. For
example, at the $50^{th}$ percentile, the mean (variance) of the
absolute prediction errors for FPCA, CP-based and Tucker-based models
are $0.12(0.02)$, $0.07(0.006)$ and $0.05(0.003)$, respectively.
A similar pattern can also be seen at the remaining prediction percentiles.
As mentioned earlier, one explanation for this phenomenon is that
by averaging the pixel intensities we break the spatio-temporal structure
of the image, which is clearly an important aspect that needs to be
considered when modeling how images evolve spatially and temporally-{}-a
key aspect that is addressed by our modeling framework.

Figure \ref{fig:er} also shows that Tucker-based regression performs
better than CP-based. The mean and variance for the Tucker-based model
are consistently lower than those of the CP-based regression model.
This difference may be attributed to the fact that the Tucker-based
model allows the tensor to have a different rank for each of the three
orders (directions). This enhances the flexibility of the regression
model. In contrast, the CP-based model requires the rank on each direction
to be equal, which may have an impact on the model's flexibility.

\section{Conclusions}

\label{sec:conclusion}

Degradation tensors such image streams and profiles often contain
rich information about the physical degradation process and can be
utilized for prognostics and predicting the RUL of functioning systems.
However, the analysis of degradation tensors is challenging due to
their high-dimensionality and complex spatial-temporal structure.
In this paper, we proposed a penalized (log)-location-scale regression
model that can utilize high dimensional tensors to predict the RUL
of systems. Our method first reduces the dimensionality of tensor
covariates by projecting them onto a low-dimensional tensor subspace
that preserves the useful information of the covariates. Next, the
projected low-dimensional covariate tensors are regressed against
TTFs via an LLS regression model. In order to further reduce the number
of parameters, the coefficient tensor is decomposed by utilizing two
tensor decompositions, CP and Tucker. The CP decomposition decomposes
the coefficient tensor as a product of low-dimensional basis matrices,
and Tucker decomposition expresses it as a product of a low-dimensional
core tensor and factor matrices. Instead of estimating the coefficient
tensor, we only estimate its corresponding core tensors and factor/basis
matrices. By doing so, the number of parameters to be estimated is
dramatically reduced. Two numerical block relaxation algorithms with
global convergence property were developed for the model estimation.
The block relaxation algorithms iteratively estimate only one block
of parameters (i.e., one factor/basis matrix or core tensor) at each
time while keeping other blocks fixed until convergences.

We evaluated the performance of our proposed methodology through numerical
studies. The results indicated that our methodology outperformed the
benchmark in terms of both prediction accuracy and precision. For
example,  the mean prediction error for CP-based tensor regression
model is 1\%, while it is 15\% for the benchmark. In addition, we
showed that the absolute mean prediction error for Tucker-based regression
model and the benchmark are 1\% and 20\%, respectively. We also
validated the effectiveness of our proposed tensor regression model
using a case study on degradation modeling of bearings in a rotating
machinery. The results indicated that both CP-based  and Tucker-based
models outperformed the benchmark in terms of  prediction accuracy
as well as precision at all life percentiles. As an example, the mean
(variance) of prediction errors at the 50th observation percentile
are 12\% (2\%), 7\% (0.6\%) and 5\% (0.3\%) for FPCA, CP-based model
and Tucker-based model, respectively. The results also indicated that
Tucker-based model achieved better prediction accuracy than the CP-based
model. This is reasonable since Tucker-based model is more flexible
as it allows different modes to have different ranks, while the CP-based
model requires all the modes have the same rank. The model developed
in this paper only works on a single computer. Development of a tensor-based
prognostics model that can run on a distributed computing system is
an important topic for future research.\bigskip{}

\appendix
%dummy comment inserted by tex2lyx to ensure that this paragraph is not empty%dummy comment inserted by tex2lyx to ensure that this paragraph is not empty\newpage{}APPENDIX

\section{Proof of Proposition \ref{prop:dim}\label{app:Proof-for-Propositiondim}}

The proof follows the proof of Lemma 1 in \citet{Li2013}. Specifically, the mode-$d$ matricization of tensor $\mathcal{S}_{i}$ and $\tilde{\mathcal{B}}$ can be expressed as \citep{Li2013}:

\[
\begin{split}
\boldsymbol{S}_{i(d)}=\boldsymbol{U}_{d}\tilde{\boldsymbol{S}}_{i(d)}\left(\boldsymbol{U}_{D}\otimes\cdots\otimes\boldsymbol{U}_{d+1}\otimes\boldsymbol{U}_{d-1}\otimes\cdots\otimes\boldsymbol{U}_{1}\right)^{\top},\\
\tilde{\boldsymbol{B}}_{(d)}=\boldsymbol{U}_{d}^\top{\boldsymbol{B}}_{(d)}\left(\boldsymbol{U}_{D}\otimes\cdots\otimes\boldsymbol{U}_{d+1}\otimes\boldsymbol{U}_{d-1}\otimes\cdots\otimes\boldsymbol{U}_{1}\right).
\end{split}
\]
\\
 Then, we have the following: 
\[
\begin{split} & \langle\mathcal{B},\mathcal{S}_{i}\rangle\\
 & =\langle\boldsymbol{B}_{(d)},\boldsymbol{S}_{i(d)}\rangle\\
 & =\langle{\boldsymbol{B}}_{(d)},\boldsymbol{U}_{d}\tilde{\boldsymbol{S}}_{i(d)}\left(\boldsymbol{U}_{D}\otimes\cdots\otimes\boldsymbol{U}_{d+1}\otimes\boldsymbol{U}_{d-1}\otimes\cdots\otimes\boldsymbol{U}_{1}\right)^{\top}\rangle\\
 & =\langle\boldsymbol{U}_{d}^\top{\boldsymbol{B}}_{(d)}\left(\boldsymbol{U}_{D}\otimes\cdots\otimes\boldsymbol{U}_{d+1}\otimes\boldsymbol{U}_{d-1}\otimes\cdots\otimes\boldsymbol{U}_{1}\right),\tilde{\boldsymbol{S}}_{i(d)}\rangle\\
 & =\langle\tilde{\boldsymbol{B}}_{(d)},\tilde{\boldsymbol{S}}_{i(d)}\rangle\\
 & =\langle\tilde{\mathcal{B}},\tilde{\mathcal{S}}_{i}\rangle
\end{split}
\]

\section{Optimization Algorithm for Problem \eqref{eq:mpca}\label{app:Optimization-mpca}\protect }

The pseudocode of the algorithm is shown in Table \ref{lb:mpca} \citep{Lu2008}

\begin{table}
\centering{} %
\begin{tabular}{|>{\raggedright}p{16cm}|}
\hline 
\textbf{Input:} A set of tensor samples $\left\{ \mathcal{S}_{i}\in R^{I_{1}\times\cdots\times I_{D}}\right\} _{i=1}^{N}$ \tabularnewline
\textbf{Output:} Low-dimensional representations $\left\{ \mathcal{\tilde{S}}_{i}\in R^{P_{1}\times\cdots\times P_{D}}\right\} _{i=1}^{N}$of
the input tensor samples with maximum variation captured \tabularnewline
\textbf{Algorithm:}\tabularnewline
\begin{tabular}{l>{\raggedright}p{15cm}}
 & \textbf{Step 1 (Preprocessing): }Center the input samples as $\left\{ \mathcal{X}_{i}=\mathcal{S}_{i}-\bar{\mathcal{S}}\right\} _{i=1}^{N}$
, where $\bar{\mathcal{S}}=\frac{1}{N}\sum_{i=1}^{N}\mathcal{S}_{i}$
is the sample mean\tabularnewline
 & \textbf{Step 2 (Initialization):} Calculate the eigen-decomposition
of $\boldsymbol{\Phi}_{(d)}^{*}=\sum_{i=1}^{N}\boldsymbol{X}_{i(d)}\boldsymbol{X}_{i(d)}^{\top}$
and set $\boldsymbol{U}_{d}$ to consist of the eigenvectors corresponding
to the most significant $P_{d}$ eigenvalues, for $d=1,\ldots D$\tabularnewline
 & \textbf{Step 3 (Local optimization):} \tabularnewline
 & %
\begin{tabular}{l>{\raggedright}p{14.5cm}}
 & (i) Calculate $\left\{ \tilde{\mathcal{X}_{i}}=\mathcal{X}_{i}\times_{1}\boldsymbol{U}_{1}^{\top}\times_{2}\boldsymbol{U}_{2}^{\top}\times\cdots\times_{D}\boldsymbol{U}_{D}^{\top}\right\} _{i=1}^{N}$\tabularnewline
 & (ii) Calculate $\Psi_{0}=\sum_{i=1}^{N}||\tilde{\mathcal{X}_{i}}||_{F}^{2}$
(the mean of $\left\{ \tilde{\mathcal{X}_{i}}\right\} _{i=1}^{N}$
is all zero since $\left\{ \mathcal{X}_{i}\right\} _{i=1}^{N}$is
centered.\tabularnewline
 & (iii) For $k=1:K$\tabularnewline
 & %
\begin{tabular}{l>{\raggedright}p{14cm}}
 & \textendash{} For $d=1:D$

\begin{tabular}{l>{\raggedright}p{13.5cm}}
 & Calculate the eigen-decomposition of $\boldsymbol{\Phi}_{(d)}$ and
set $\boldsymbol{U}_{d}$ to consist of the eigenvectors corresponding
to the most significant $P_{d}$ eigenvalues, for $d=1,\ldots D$,
where $\boldsymbol{\Phi}_{(d)}=\sum_{i=1}^{N}\left(\boldsymbol{S}_{i(d)}-\bar{\boldsymbol{S}}_{(d)}\right)\cdot\boldsymbol{A}_{d}\cdot\boldsymbol{A}_{d}^{\top}\cdot\left(\boldsymbol{S}_{i(d)}-\bar{\boldsymbol{S}}_{(d)}\right)^{\top}$
and $\boldsymbol{A}_{d}=\left(\boldsymbol{U}_{d+1}\otimes\boldsymbol{U}_{d+2}\otimes\cdots\otimes\boldsymbol{U}_{D}\otimes\boldsymbol{U}_{1}\otimes\boldsymbol{U}_{2}\otimes\cdots\otimes\boldsymbol{U}_{d-1}\right)$\tabularnewline
\end{tabular}\tabularnewline
 & \textendash{} Calculate $\left\{ \tilde{\mathcal{S}_{i}}\right\} _{i=1}^{N}$
and $\Psi_{k}$\tabularnewline
 & \textendash{} If $\Psi_{k}-\Psi_{k}<\eta$, break and go to Step 4.\tabularnewline
\end{tabular}\tabularnewline
\end{tabular}\tabularnewline
 & \textbf{Step 4 (Projection):} The feature tensor after projection
is obtained as $\left\{ \tilde{\mathcal{S}_{i}}=\mathcal{S}_{i}\times_{1}\boldsymbol{U}_{1}^{\top}\times_{2}\boldsymbol{U}_{2}^{\top}\times\cdots\times_{D}\boldsymbol{U}_{D}^{\top}\right\} _{i=1}^{N}$\tabularnewline
\end{tabular}\tabularnewline
\tabularnewline
\hline 
\end{tabular}\caption{Pseudocode implementation of the MPCA algorithm \citep{Lu2008}.}
\label{lb:mpca} 
\end{table}

\section{Proof of Proposition \ref{prop:cp}\label{app:Proof-for-Propositioncp}}

Based on the CP decomposition, tensor $\mathcal{B}$ has the following
properties \citep{Li2013}:

$\begin{array}{cc}
 & vec({\mathcal{B}})=(\boldsymbol{B}_{D}\odot\cdots\odot\boldsymbol{B}_{1})\boldsymbol{1}_{R},\\
 & \boldsymbol{B}_{(d)}=\boldsymbol{B}_{d}(\boldsymbol{B}_{D}\odot\cdots\odot\boldsymbol{B}_{d+1}\odot\boldsymbol{B}_{d-1}\odot\cdots\odot\boldsymbol{B}_{1})^{\top}
\end{array}$

Recall the optimization problem is

\[
\argmax_{\boldsymbol{\theta}}\left\{ -N\log\sigma+\sum_{i=1}^{N}\log f\left(\frac{y_{i}-\alpha-\left\langle (\tilde{\boldsymbol{B}}_{D}\odot\cdots\odot\tilde{\boldsymbol{B}}_{1})\boldsymbol{1}_{R},{vec}(\tilde{\mathcal{S}}_{i})\right\rangle }{\sigma}\right)-\sum_{d=1}^{D}r\left(\tilde{\boldsymbol{B}}_{d}\right)\right\} 
\]

Given $\left\{ \tilde{\boldsymbol{B}}_{1},\ldots,\tilde{\boldsymbol{B}}_{d-1},\tilde{\boldsymbol{B}}_{d+1},\ldots,\tilde{\boldsymbol{B}}_{D}\right\} $,the
inner product in the optimization is

$\begin{array}{cl}
 & \langle(\boldsymbol{B}_{D}\odot\cdots\odot\boldsymbol{B}_{1})\boldsymbol{1}_{R},{vec}(\tilde{\mathcal{S}}_{i})\rangle\\
= & \langle{vec}(\tilde{\mathcal{B}}),{vec}(\tilde{\mathcal{S}}_{i})\rangle\\
= & \langle\tilde{\mathcal{B}},\tilde{\mathcal{S}}_{i}\rangle\\
= & \langle\tilde{\boldsymbol{B}}_{(d)},\tilde{\boldsymbol{S}}_{i(d)}\rangle\\
= & \langle\boldsymbol{B}_{d}(\boldsymbol{B}_{D}\odot\cdots\odot\boldsymbol{B}_{d+1}\odot\boldsymbol{B}_{d-1}\odot\cdots\odot\boldsymbol{B}_{1})^{\top},\tilde{\boldsymbol{S}}_{i(d)}\rangle\\
= & \langle\boldsymbol{B}_{d},\tilde{\boldsymbol{S}}_{i(d)}(\boldsymbol{B}_{D}\odot\cdots\odot\boldsymbol{B}_{d+1}\odot\boldsymbol{B}_{d-1}\odot\cdots\odot\boldsymbol{B}_{1})\rangle\\
= & \langle\boldsymbol{B}_{d},\boldsymbol{X}_{d,i}\rangle
\end{array}$

where $\boldsymbol{X}_{d,i}=\tilde{\boldsymbol{S}}_{i(d)}(\boldsymbol{B}_{D}\odot\cdots\odot\boldsymbol{B}_{d+1}\odot\boldsymbol{B}_{d-1}\odot\cdots\odot\boldsymbol{B}_{1})$.
Therefore, the optimization problem can be re-expressed as

\[
\argmax_{\boldsymbol{\theta}}\left\{ -N\log\sigma+\sum_{i=1}^{N}\log f\left(\frac{y_{i}-\alpha-\langle\boldsymbol{B}_{d},\boldsymbol{X}_{d,i}\rangle}{\sigma}\right)-\sum_{d=1}^{D}r\left(\tilde{\boldsymbol{B}}_{d}\right)\right\} 
\]

\section{Invariant Property of Optimization Problem \eqref{eq:bd4}\label{app:invariant}}

Recall optimization problem \eqref{eq:bd4}

\[
\argmax_{\boldsymbol{\tilde{B}}_{d},\sigma,\alpha}\left\{ -N\ln\sigma+\sum_{i=1}^{N}\ln f\left(\frac{y_{i}-\alpha-\left\langle \tilde{\boldsymbol{B}}_{d},\boldsymbol{X}_{d,i}\right\rangle }{\sigma}\right)-r(\frac{\tilde{\boldsymbol{B}}_{d}}{\sigma})\right\} 
\]

Consider the transformation $y_{i}'=by_{i},\alpha'=b\alpha,\tilde{\boldsymbol{B}}_{d}'=b\tilde{\boldsymbol{B}_{d}},\sigma'=b\sigma$
where $b>0$, we have

$\begin{array}{cl}
 & \argmax_{\boldsymbol{\tilde{B}}_{d}',\sigma',\alpha'}\left\{ -N\ln\sigma'+\sum_{i=1}^{N}\ln f\left(\frac{y_{i}'-\alpha'-\left\langle \tilde{\boldsymbol{B}}_{d}',\boldsymbol{X}_{d,i}\right\rangle }{\sigma}\right)-r(\frac{\tilde{\boldsymbol{B}}_{d}'}{\sigma'})\right\} \\
\iff & \argmax_{\boldsymbol{\tilde{B}}_{d},\sigma,\alpha}\left\{ -N\ln\left(b\sigma\right)+\sum_{i=1}^{N}\ln f\left(\frac{by_{i}-b\alpha-\left\langle b\tilde{\boldsymbol{B}}_{d},\boldsymbol{X}_{d,i}\right\rangle }{b\sigma}\right)-r(\frac{b\tilde{\boldsymbol{B}}_{d}}{b\sigma})\right\} \\
\iff & \argmax_{\boldsymbol{\tilde{B}}_{d},\sigma,\alpha}\left\{ -N\ln\left(b\sigma\right)+\sum_{i=1}^{N}\ln f\left(\frac{by_{i}-b\alpha-b\left\langle \tilde{\boldsymbol{B}}_{d},\boldsymbol{X}_{d,i}\right\rangle }{b\sigma}\right)-r(\frac{b\tilde{\boldsymbol{B}}_{d}}{b\sigma})\right\} \\
\iff & \argmax_{\boldsymbol{\tilde{B}}_{d},\sigma,\alpha}\left\{ -N\ln b-N\ln\sigma+\sum_{i=1}^{N}\ln f\left(\frac{y_{i}-\alpha-\left\langle \tilde{\boldsymbol{B}}_{d},\boldsymbol{X}_{d,i}\right\rangle }{\sigma}\right)-r(\frac{\tilde{\boldsymbol{B}}_{d}}{\sigma})\right\} \\
\iff & \argmax_{\boldsymbol{\tilde{B}}_{d},\sigma,\alpha}\left\{ -N\ln\sigma+\sum_{i=1}^{N}\ln f\left(\frac{y_{i}-\alpha-\left\langle \tilde{\boldsymbol{B}}_{d},\boldsymbol{X}_{d,i}\right\rangle }{\sigma}\right)-r(\frac{\tilde{\boldsymbol{B}}_{d}}{\sigma})\right\} 
\end{array}$

\section{Proof of Proposition \ref{prop:tucker1}\label{app:Proof-for-Propositiontucker}}

Based on the Tucker decomposition, tensor $\mathcal{B}$ has the following
properties \citep{Li2013}:

$\begin{array}{cc}
 & \boldsymbol{B}_{(d)}=\boldsymbol{B}_{d}\boldsymbol{G}_{(d)}(\boldsymbol{B}_{D}\otimes\cdots\otimes\boldsymbol{B}_{d+1}\otimes\boldsymbol{B}_{d-1}\otimes\cdots\otimes\boldsymbol{B}_{1})^{\top},\\
 & vec({\mathcal{B}})=(\boldsymbol{B}_{D}\otimes\cdots\otimes\boldsymbol{B}_{1})vec(\mathcal{G}).
\end{array}$

Recall the optimization problem is
\begin{align}
  & \argmax_{\boldsymbol{\theta}}\left\{ -N\ln\sigma+\sum_{i=1}^{N}\ln f\left(\frac{y_{i}-\alpha-\left\langle \mathcal{\tilde{G}}\times_{1}\tilde{\boldsymbol{B}}_{1}\times_{2}\tilde{\boldsymbol{B}}_{2}\times_{3}\cdots\times_{D}\tilde{\boldsymbol{B}}_{D},\tilde{\mathcal{S}}_{i}\right\rangle }{\sigma}\right)\right.\nonumber \\\nonumber
 & \left.-r(\mathcal{\mathcal{\tilde{G}}})-\sum_{d=1}^{D}r\left(\tilde{\boldsymbol{B}}_{d}\right)\right\},
\end{align}

Given $\left\{ \tilde{\boldsymbol{B}}_{1},\tilde{\boldsymbol{B}}_{2}\ldots,\tilde{\boldsymbol{B}}_{D}\right\} $,
the inner product in the optimization can be expressed as

$\begin{array}{cl}
 & \langle\mathcal{\tilde{G}}\times_{1}\tilde{\boldsymbol{B}}_{1}\times_{2}\tilde{\boldsymbol{B}}_{2}\times_{3}\cdots\times_{D}\tilde{\boldsymbol{B}}_{D},\tilde{\mathcal{S}}_{i}\rangle\\
= & \langle\tilde{\mathcal{B}},\tilde{\mathcal{S}}_{i}\rangle\\
= & \langle vec(\tilde{\mathcal{B}}),vec(\tilde{\mathcal{S}}_{i})\rangle\\
= & \langle(\tilde{\boldsymbol{B}}_{D}\otimes\cdots\otimes\tilde{\boldsymbol{B}}_{1})vec(\mathcal{\mathcal{\mathcal{\tilde{G}}}}),vec(\tilde{\mathcal{S}}_{i})\rangle\\
= & \langle vec(\mathcal{\mathcal{\mathcal{\tilde{G}}}}),(\tilde{\boldsymbol{B}}_{D}\otimes\cdots\otimes\tilde{\boldsymbol{B}}_{1})^{\top}vec(\tilde{\mathcal{S}}_{i})\rangle
\end{array}$

Therefore, the optimization problem can be re-expressed as

\[
\argmax_{\mathcal{\mathcal{\mathcal{\mathcal{\tilde{G}}}}}}\left\{ -N\ln\sigma+\sum_{i=1}^{N}\ln f\left(\frac{y_{i}-\alpha-\left\langle vec(\mathcal{\mathcal{\mathcal{\mathcal{\tilde{G}}}}}),(\tilde{\boldsymbol{B}}_{D}\otimes\cdots\otimes\tilde{\boldsymbol{B}}_{1})^{\top}vec(\tilde{\mathcal{S}}_{i})\right\rangle }{\sigma}\right)-r(\mathcal{\mathcal{\mathcal{\mathcal{\tilde{G}}}}})\right\} ,
\]

\section{Proof for Proposition \ref{prop:tucker2}\label{app:Proof-for-Propositiontucker2}}

Recall the optimization problem is

\begin{align}
  & \argmax_{\boldsymbol{\theta}}\left\{ -N\ln\sigma+\sum_{i=1}^{N}\ln f\left(\frac{y_{i}-\alpha-\left\langle \mathcal{\tilde{G}}\times_{1}\tilde{\boldsymbol{B}}_{1}\times_{2}\tilde{\boldsymbol{B}}_{2}\times_{3}\cdots\times_{D}\tilde{\boldsymbol{B}}_{D},\tilde{\mathcal{S}}_{i}\right\rangle }{\sigma}\right)\right.\nonumber \\\nonumber
 & \left.-r(\mathcal{\mathcal{\tilde{G}}})-\sum_{d=1}^{D}r\left(\tilde{\boldsymbol{B}}_{d}\right)\right\},
\end{align}

Given $\mathcal{\tilde{G}}$ and $\left\{ \tilde{\boldsymbol{B}}_{1},\ldots,\tilde{\boldsymbol{B}}_{d-1},\tilde{\boldsymbol{B}}_{d+1},\ldots,\tilde{\boldsymbol{B}}_{D}\right\} $,
the inner product in the optimization can be expressed as

$\begin{array}{ll}
 & \langle\mathcal{G}\times_{1}\boldsymbol{B}_{1}\times_{2}\boldsymbol{B}_{2}\times_{3}\cdots\times_{D}\boldsymbol{B}_{D},\tilde{\mathcal{S}}_{i}\rangle\\
= & \langle\tilde{\mathcal{B}},\tilde{\mathcal{S}}_{i}\rangle\\
= & \langle\tilde{\boldsymbol{B}}_{(d)},\tilde{\boldsymbol{S}}_{i(d)}\rangle\\
= & \langle\boldsymbol{B}_{d}\boldsymbol{G}_{(d)}(\boldsymbol{B}_{D}\otimes\cdots\otimes\boldsymbol{B}_{d+1}\otimes\boldsymbol{B}_{d-1}\otimes\cdots\otimes\boldsymbol{B}_{1})^{\top},\tilde{\boldsymbol{S}}_{i(d)}\rangle\\
= & \langle\boldsymbol{B}_{d},\tilde{\boldsymbol{S}}_{i(d)}(\boldsymbol{B}_{D}\otimes\cdots\otimes\boldsymbol{B}_{d+1}\otimes\boldsymbol{B}_{d-1}\otimes\cdots\otimes\boldsymbol{B}_{1})\boldsymbol{G}_{(d)}^{\top}\rangle\\
= & \langle\boldsymbol{B}_{d},\boldsymbol{X}_{d,i}\rangle
\end{array}$

where $\boldsymbol{X}_{d,i}=\tilde{\boldsymbol{S}}_{i(d)}(\boldsymbol{B}_{D}\otimes\cdots\otimes\boldsymbol{B}_{d+1}\otimes\boldsymbol{B}_{d-1}\otimes\cdots\otimes\boldsymbol{B}_{1})\boldsymbol{G}_{(d)}^{\top}$.
Therefore, the optimization problem can be re-expressed as

\[
\argmax_{\boldsymbol{\theta}}\left\{ -N\ln\sigma+\sum_{i=1}^{N}\ln f\left(\frac{y_{i}-\alpha-\langle\boldsymbol{B}_{d},\boldsymbol{X}_{d,i}\rangle}{\sigma}\right)\left.-r(\mathcal{\mathcal{\tilde{G}}})-\sum_{d=1}^{D}r\left(\tilde{\boldsymbol{B}}_{d}\right)\right\} .\right.
\]

{}

\end{document}